\documentclass[a4paper,11pt]{article}
\usepackage{jheppub} 
\usepackage{lineno}
\usepackage{amsmath,amssymb,amsfonts,bm}
\usepackage{graphicx}
\usepackage{xcolor}
\usepackage{bbold}
\usepackage{enumitem}
\usepackage{graphicx}
\usepackage{dcolumn}
\usepackage{epstopdf}
 \usepackage[thinlines,thiklines]{easybmat}
\usepackage{epsfig}
\usepackage[caption=false]{subfig}
\usepackage{url}
\usepackage{hyperref}
\usepackage{verbatim}
\usepackage[export]{adjustbox}
\usepackage{scalerel}
\usepackage{braket}
\usepackage{ytableau}
\usepackage{amsmath}

\usepackage{tikz}
\usepackage{ulem}

\newcommand{\Texp}{\mathcal{T}\!\exp}

\newcommand{\Tr}{\operatorname{Tr}}

\def\ii{\textbf{i}}
\def\jj{\textbf{j}}

\title{\boldmath Frame potential of Brownian SYK model of Majorana and Dirac fermions}

\author[1]{Anastasiia Tiutiakina}
\affiliation{Laboratoire de Physique Th\'eorique et Mod\'elisation, CNRS UMR 8089,
	CY Cergy Paris Universit\'e, 95302 Cergy-Pontoise Cedex, France}

\author[1]{Andrea De Luca}

\author[1]{Jacopo De Nardis}

\emailAdd{anastasiia.tiutiakina@cyu.fr}

\abstract{We consider the Brownian SYK, i.e. a system of $N$ Majorana (Dirac) fermions with a white-noise $q$-body interaction term. We focus on the dynamics of the Frame potentials, a measure of the scrambling and chaos, given by the moments of the overlap between two independent realisations of the model. By means of a Keldysh path-integral formalism, we compute its early and late-time value. We show that, for $q>2$, the late time path integral saddle point correctly reproduces the saturation to the value of the Haar frame potential. On the contrary, for $q=2$, the model is quadratic and consistently we observe saturation to the Haar value in the restricted space of Gaussian states (Gaussian Haar). The latter is characterised by larger system size corrections that we correctly capture by counting the Goldstone modes of the Keldysh saddle point. Finally, in the case of Dirac fermions, we highlight and resolve the role of the global $U(1)$ symmetry.  }

\begin{document}
\maketitle
\flushbottom

\section{Introduction and main results}
\label{sec:intro}
The phenomenon of information scrambling in quantum many-body systems has recently gained significant attention due to its deep connections with holography, black hole physics, and quantum information \cite{Hayden2007, Shenker2014, Shenker2014-2, Maldacena2016, PRXQuantum.2.010306, Almheiri2020, PhysRevX.9.031048,jian2022linear}. While some quantum scrambler quantifiers are inspired by the classical theory of chaos, as is the case for the OTOCs (out-of-time-order correlations), a slightly different way to look at information scrambling is via $k$-design realisations. Namely, one asks when and how the unitary time evolution from a pure initial state can generate an ensemble of random quantum states (under a given definition of such an ensemble) which, up to the $k$-th moment, is uniformly (Haar) distributed  ~\cite{Gross2007, Roberts2017,Hearth2023, PhysRevX.13.011043}.
Such a question is directly related to the emergence of eigenstate thermalization (ETH) in closed quantum systems~\cite{PhysRevE.50.888,DAlessio2016,PhysRevLett.129.170603,PhysRevE.57.7313,PhysRevA.43.2046}. More recently, a related problem has been posed in the context of deterministic Hamiltonian systems undergoing final read-out measurements: the randomness induced by quantum measurements can be approximated with different accuracy by a Haar random ensembles, a feature recently dubbed \textit{deep thermalization}~\cite{Lucas2022, Hearth2023, PRXQuantum.4.010311, Choi2023}. Generic chaotic quantum systems are indeed expected to thermalise at a late time, i.e. the reduced density matrices obtained by tracing out most of the degrees of freedom are expected to converge to the standard Gibbs statistical density matrix, which reduces to the identity matrix in the absence of conserved quantities. However, the presence of different sources of stochasticity (e.g. noise or measurements), raises the question of whether, beyond the convergence of the stochastic average, individual realisation (or trajectories) of the system are exploring uniformly the accessible Hilbert space of pure states. However, the study of strongly interacting systems, in particular with the additional ingredient of stochasticity, is generally analytically (and also numerically) very challenging, except for a few specific systems made of random matrices or dual unitary quantum circuits \cite{Ho2022, Ho2023, Ippoliti2022}.
For this reason, it is important to develop an intuition on some sufficiently generic but tractable cases. Among these, a prominent example is the Brownian SYK model \cite{PhysRevLett.70.3339, 1806.06840}, describing $L$ Majorana (or Dirac) fermions with all-to-all interactions (as for the standard SYK), with a time-dependent white-noise interaction coupling. The model is well known to serve as a paradigmatic model for quantum scrambling and chaos \cite{Jian2021, 1806.06840, Snderhauf2019, PhysRevE.78.021106, PhysRevX.9.031048, PhysRevB.100.064305}.

In this paper, we focus on investigating the degree of mixing in the time evolution of the Brownian SYK model. To address this question, we employ a measure of mixing called the Frame potential. Specifically, we consider two identical copies of an initial density matrix $\rho_0$ and compute the overlap between the time evolutions of these copies under two independent realisations of the noise. The Frame potential, denoted as $F^{(k)}(T)$, is given by the expression:
\begin{equation}\label{fr0}
F^{(k)}(T)= \left< {\rm Tr} \left(U_1(T) \rho_0 U_1^{\dagger}(T) U_2(T) \rho_0 U_2^{\dagger}(T) \right)^k \right>,
\end{equation}
Here, $U_1$ and $U_2$ represent the (noisy) evolution operators of two independent Brownian SYK systems with randomly chosen couplings and $F^{(k)}$ equals the $k$-th moment (averaged over both noise realizations) of their overlap. In the case of sufficiently mixing time evolutions and starting from a reference pure state $\rho_0 = \ket{\Psi}\bra{\Psi}$, the Frame potential exhibits an exponential decay from its initial value one. This decay manifests the decreasing probability that the two density matrices remain close as time progresses. Eventually, at infinite time $T \to \infty$, it is expected that the density matrix $\rho(T)$ fully explores the entire Hilbert space in a uniform manner, modulo the constraints imposed by global symmetries as $U(1)$, implying that its statistics can be effectively obtained by replacing the time-evolution operator with a Haar distributed unitary operator in the Hilbert space $\mathcal{H}$ as 
\begin{equation}\label{eq:latetimeconv}
\rho(T) \to U^\dagger \rho_0 U, \qquad U \sim \mbox{Haar}(\mathcal{H}).
\end{equation}
Consequently, the Frame potentials for all $k$'s are expected to converge at large times to their Haar-averaged value, i.e. where both $U_1(T)$ and $U_2(T)$ are replaced by two independently drawn Haar-distributed unitary matrices. We denote this limit case as $F_{\rm Haar}^{(k)}$ and it is easy to prove (see for instance \cite{Ippoliti2022}) that it corresponds to the minimum overall possible distributions of unitary operators on $\mathcal{H}$. As detailed in this paper, this convergence indeed occurs 
for any $q > 2$ (with interactions coupling more than 2 fermions), thus generalising the results of~\cite{jian2022linear}. An exception arises in the Gaussian integrable case $q=2$, with exactly two-fermions interaction~\cite{10.21468/SciPostPhys.12.1.042, Bernard2021, swann2023spacetime}. In this case, the evolution only exhibits mixing behaviour within the space of Gaussian states. There, we can introduce a Gaussian Haar (gHaar)~\cite{PhysRevE.104.014146} measure and demonstrate that the Frame potentials converge for all $k$ to the latter at late times. Intriguingly, the value of the gHaar-integrated Frame potential is equivalent to the generic case in the limit of a large fermion number $L$, with logarithmic corrections in $L$. As demonstrated in this work, these corrections arise from fluctuations of Goldstone modes around the late-time saddle point, which are only present in the Gaussian case where both the Keldysh action and the saddle point's symmetries are generated by continuous groups.
Moreover, we show that in the case of Dirac SYK  the global $U(1)$ symmetry leads to a late-time convergence \eqref{eq:latetimeconv} in each sector of a given fermionic charge, giving a radically different behaviour between the non-Gaussian and the Gaussian case. In both cases, the Frame potential converges at late times to a larger value compared to its Majorana equivalent, as the result of the global constraints given by the $U(1)$ symmetry. 

\begin{figure}
    \centering
    \includegraphics[width=1 \textwidth]{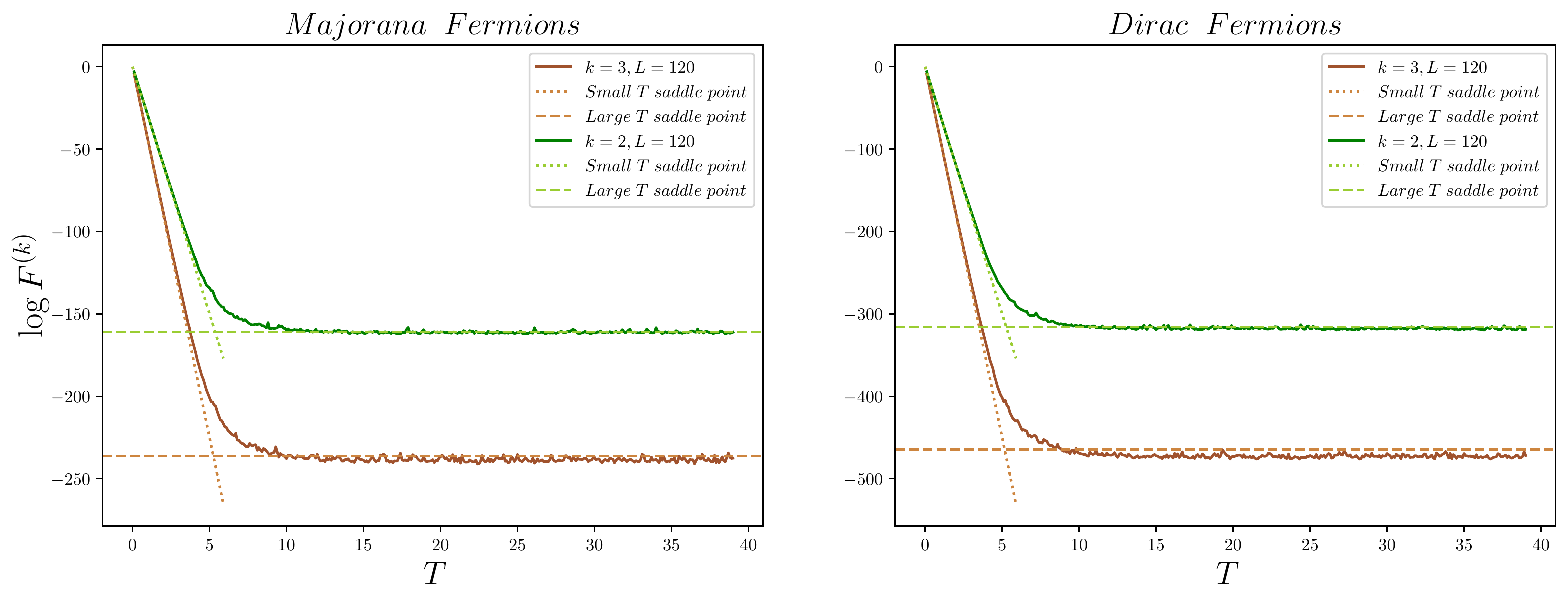}
    \caption{The rescaled Frame potential with $q=2$ in the logarithmic scale, exact numerical simulations compared with the theoretical predictions. The number of fermions in both cases is $L=120$. Brown colour corresponds to the number of replicas $k=3$, green colour corresponds to $k=2$. The dotted line is the theoretical prediction of the Keldysh calculation at early times, and the dashed line is the prediction for the late time behaviour.
   From the numerical data, we can see that the time of the saturation can be estimated directly from the theoretical prediction, modulo finite size corrections (which are larger for larger $k$ as expected). The intersection between the short-time saddle point (dotted line) and the long-time saddle point (dashed line)  gives the estimation for the saturation time. In the case $q=2$:    $
    T^{q=2}_{\rm Dirac}=4\left(2\log(2)- \frac{k}{L} \log(L) \right)+O\left(\frac{1}{L}\right),~~T^{q=2}_{\rm Majorana}=4\left(2\log(2)- \frac{k-1}{L} \log(L) \right)+O\left(\frac{1}{L}\right);$
as compared to the case $q>2$, where we have $ T^{q>2}_{\rm Dirac}=T^{q>2}_{\rm Majorana}=q^2\cdot2\log(2)+O\left(\frac{1}{L}\right)$.}
    \label{fig:num}
\end{figure}
Summarizing our results here, we find the following: 
\begin{itemize}
    \item the Frame potential decays at short times as   \begin{align}
        F^{(k)}_{\rm Haar} &  = e^{-\frac{LTk}{2q^2}} \quad \quad {\rm Majorana},  \nonumber \\
         F^{(k)}_{\rm Haar} &  = e^{-\frac{LTk}{q^2}} \quad \quad {\rm Dirac}. \nonumber
    \end{align}
    \item For $q>2$, given the dimension of the Hilbert space $\mathcal{N}$, the Frame potential saturates at large times to the values 
    \begin{align}
        F^{(k)}_{\rm Haar} &  = \frac{k!}{\mathcal{N}^{2k}} \quad \quad {\rm Majorana},  \nonumber \\
         F^{(k)}_{\rm Haar} &  = \frac{L^k k!}{\mathcal{N}^{2k}} \quad \quad {\rm Dirac}. \nonumber
    \end{align}
    \item  For $q=2$, including the powers in $L$ due to massless fluctuation around the saddle point, the Frame potential saturates at large times to the values, whose leading terms in $L$ are given by 
    \begin{align}
        F^{(k)}_{\rm gHaar} &  = \frac{c_k L^{k(k-1)/2}}{\mathcal{N}^{2k}}+ \ldots \quad \quad {\rm Majorana}, \nonumber \\
         F^{(k)}_{\rm gHaar} &  = \frac{\tilde{c}_k L^{k^2}}{\mathcal{N}^{2k}} + \ldots \quad \quad {\rm Dirac}, \nonumber
    \end{align}
\end{itemize}
where $c_k={2^{1-\frac{(k-1)(k-2)}{2}}} \prod_{i=0}^{k-1}1/(2i-1)!! $ and $\tilde{c}_k={{\rm sf}^2(k-1)}/{{\rm sf}(2k-1)}$, with ${\rm sf}(n)=1!2!...n!$.

The structure of this paper is as follows. In Section \ref{sec2}, we derive the Frame potential for the Brownian Majorana SYK system using the framework of the thermofield double state. We outline the Keldysh path integral approach, which facilitates averaging over the Brownian noise, and obtain an effective action for the problem. Moving to Section \ref{sec3}, we explore the Swinger-Dyson equations by performing a variation of the action. Initially, we consider a naive saddle point solution that does not mix different Keldysh contours and replicas. The resulting Frame potential exhibits exponential decay, as depicted by the dotted lines in Fig. (\ref{fig:num}). Subsequently, we investigate a non-trivial solution in the large time limit, wherein different replicas and contours are mixed. Notably, this solution exhibits distinct group symmetries for cases with $q>2$ and $q=2$. The continuous symmetry group characterizes the $q=2$ case, while discrete symmetries are present for $q>2$. We then proceed to analyse the fluctuations around the saddle points. For the $q>2$ case, these fluctuations trivially cancel with the normalization. However, in the $q=2$ case, fluctuations governed by the Goldstone theorem lead to massless modes in the action. Consequently, the integration over these fluctuations deviates from the normalization, resulting in polynomial corrections (in system size $L$) to the Frame potential. During the evolution, the latter reaches a saturation value exponentially small with respect to the system size. This behaviour is illustrated by the dashed lines in Fig.(\ref{fig:num}). In Section \ref{sec4}, we extend our analysis to the complex, Dirac, Brownian SYK model. Here, we show how to properly take into account the global $U(1)$ symmetry in the late-time Keldysh saddle point solution. Finally, in Section \ref{sec5}, we conclude the paper by summarizing our findings. We also discuss potential future directions and prospects for further exploration within this research framework.

\section{Brownian SYK model and Keldysh path integral representation of the Frame potential}\label{sec2}

We consider the time evolution generated by the time-dependent Hamiltonian
\begin{equation}\label{sykh}
    H(h,t)=i^{\frac{q}{2}} \sum_{1 \leq i_1<i_2<..<i_q \leq L} h_{i_1...i_q}(t) \hat \chi_{i_1}...\hat \chi_{i_q},
\end{equation}
where $\hat \chi_i$ are $L$ Majorana fermions  and anticommutation relations are applied $\{\hat \chi_i, \hat \chi_j\}=\delta_{i j}$, here $q$ is an even integer constant. This model has close analogies with the celebrated SYK~\cite{Sachdev:1992fk,Gu:2019jub,1806.06840}, but with the important difference that $h_{i_1\cdots i_q}$ are not constant in time here. On the contrary, we take them to follow a white noise distribution,
\begin{equation} \label{av}
    \left<h_{\ii}(t_1) h_{\jj}(t_2) \right>=\frac{2^{q-1}q!}{q^2 L^{q-1}  } \delta_{\ii, \jj}\delta(t_1-t_2)=\sigma^2 \delta_{\ii, \jj}\delta(t_1-t_2),
\end{equation}
where we denote the collective set of indices $\ii = (i_1,\ldots, i_q)$ (similarly for $\jj$) and set $\delta_{\ii, \jj} = \delta_{i_1, j_1} \ldots \delta_{i_q, j_q}$. For this reason, the current model is named Brownian SYK.
{We shall here assume $L$ to be an even integer, such that the operators $\hat \chi_i$ can be embedded into a $\mathcal{N} = 2^{L/2}$ dimensional Hilbert space.}
We are interested in studying the scrambling dynamics induced by the time evolution \eqref{sykh}. In order to do so, we focus on a specific initial condition, known as \textit{thermofield double} $\left(\left|{\rm TFD}\right>\right)$ \cite{Gu2017SpreadOE,1911.11977,Almheiri2020}, which has already been employed in the context of Brownian SYK in \cite{Jian2021}. In practice, we consider two copies of the system, prepared in a maximally entangled state. In the following, we shall address the two copies as left (L) and right (R) halves. The selection of the $\left|{\rm TFD}\right>$ state is not uniquely determined and relies on the choice of a basis. However, different definitions yield states that are connected through local unitary transformations.  To provide a concrete illustration, we offer an explicit characterization of $\left|{\rm TFD}\right> $  in the context of the SYK model. Within this framework, the Majorana fermion operators are denoted as $\hat \chi_{j,L}$ and $\hat \chi_{j,R}$, where $j$ takes values from $1$ to $L$. (It should be noted that $L$ must be an even number to ensure a well-defined Hilbert space). A convenient specification for the state can be formulated using Dirac fermions
\begin{equation}\label{tfdferm}
    \left(\hat c_{j,L}-\hat c^{\dagger}_{j,R}\right)\left|{\rm TFD} \right>=0~~\text{and}~~\left(\hat c^{\dagger}_{j,L}-\hat c_{j,R}\right)\left|{\rm TFD}\right>=0,
\end{equation}
which can be expressed in terms of Majorana fermions as
\begin{equation}
    \hat c_{j,L}=\frac{1}{\sqrt{2}}\left(\hat \chi_{2 j-1,L}+i \hat \chi_{2j,L} \right)~~~\text{and}~~~\hat c_{j,R}=\frac{1}{\sqrt{2}}\left(\hat \chi_{2 j,R}-i \hat \chi_{2j-1,R} \right).
\end{equation}
Then we have the standard definition in terms of Majorana fermions:
\begin{equation}
    \left(\hat \chi_{j,L(R)}+i \hat \chi_{j,L(R)}\right)\left|{\rm TFD}\right>=0,~~~ \forall j=1,...,L.
\end{equation}
From the eq.~(\ref{tfdferm}) we see that the ${\rm TFD}$ state in the eigenbasis of $n_{j,L(R)}=\hat c^{\dagger}_{j,L(R)} \hat c_{j,L(R)}$ basis can be expressed as a product of $\left|0\right>_L\left|0\right>_R+\left|1\right>_L\left|1\right>_R$ on each site. More explicitly, we consider the initial state:

\begin{equation}\label{instate}
    \left|{\rm TFD} \right> = \frac{1}{\sqrt{\mathcal{N}}}\sum_{j=1}^{\mathcal{N}} \ket{j} \ket{j},~~~~~\rho_0 = \left|{\rm TFD} \right> \left<{\rm TFD} \right|,
\end{equation}where each $\left|j \right>$ represents a possible string of zeros and ones. Here $\mathcal{N}$ is the dimension of the Hilbert space. This state has the following property, 
\begin{equation} \label{tfdprop}
    \Tr^{(2)}[A \otimes B \left|{\rm TFD} \right> \left<{\rm TFD} \right| C \otimes D] = 
    \frac{1}{\mathcal{N}} \Tr[A B^t D^t C],
\end{equation}
where the trace on the left is on the doubled Hilbert space, while the one on the right is on the single copy. This identity holds for arbitrary operators $A,B / C,D$ acting respectively on the first/second copy of the Hilbert space. 
Except for the initial entanglement, the two halves evolve according to two uncoupled unitary operators,
\begin{equation}
    \left|{\rm TFD} \right>(t) = U_{\boldsymbol{h}^L}(t) \otimes U_{\boldsymbol{h}^R}(t) \left|{\rm TFD} \right>, \qquad \rho(t) = \left|{\rm TFD}(t) \right>\left< {\rm TFD}(t)\right|,
\end{equation}
 where $U_{\boldsymbol{h}^L}$ and $U_{\boldsymbol{h}^R}$ are the unitary time evolution operators acting on the two halves. We assume that the time evolution in each half is generated by an independent realisation of the Brownian SYK
 \begin{equation}
     U_{\boldsymbol{h}^\sigma}(t + dt) = e^{-i H(\boldsymbol{h^\sigma},t) dt} U_{\boldsymbol{h}^\sigma}(t),
 \end{equation}
 where the subscript $\sigma = L,R$ labels the corresponding half 
 and $H_L(t), H_R(t)$ have the form of eq.~\eqref{sykh} in terms of two sets of Majoranas $\hat\chi_{i,\sigma}$ and independently generated white noises $h^{\sigma}_{\ii}(t)$. 
Starting with this initial state, we calculate the $k$-th moments of the Frame potential averaged over Hamiltonian realisations with measure $d\eta(\boldsymbol{h})$:
\begin{equation}\label{fr}
 F^{(k)}(T)=\int  d \eta \left(\boldsymbol{h}_1 \right)d \eta \left(\boldsymbol{h}_2\right) {\rm Tr} \left(U_{\boldsymbol{h_1}}(T) \rho_0 U^{\dagger}_{\boldsymbol{h}_1}(T)  U_{\boldsymbol{h_2}}(T) \rho_0 U^{\dagger}_{\boldsymbol{h}_2}(T) \right)^k,
\end{equation}
here, $U_{\boldsymbol{h}_1}(T)=U_{\boldsymbol{h}^L_1}(T)\otimes U^t_{\boldsymbol{h}^R_1}(T)$, (and the same for $\boldsymbol{h}^{L(R)}_2$), where the couplings $\boldsymbol{h}^{L(R)}_{1(2)}$ are independent random variables with the distribution $d \eta\left(\boldsymbol{h}\right)=e^{-\frac{\boldsymbol{h}^2}{2\sigma^2}}d \boldsymbol{h}$ and variance defined in eq.~(\ref{av}). The initial condition $\rho_0$ is the double thermofield state that we discussed earlier. Then, the Frame potential takes the form, for generic $k$, 
\begin{equation}\label{framep}
    F^{(k)}(T)=\frac{1}{\mathcal{N}^{2k}}\int d\eta(\boldsymbol{h}) \left|{\rm Tr}[U_{\boldsymbol{h}}(T)]\right|^{2k},
\end{equation}
where the evolution operator $U_{\boldsymbol{h}}(T)$ is now only acting on a single copy $R$ (or $L$) and where $\mathcal{N}$ is the dimension of its Hilbert space, coming from the definition of the initial state eq.~(\ref{instate}). In order to get to this expression, we used the property eq.~(\ref{tfdprop}) together with $U_{\boldsymbol{h}}=U_{\boldsymbol{h}^L_1}(t) U_{\boldsymbol{h}^R_1}(t) U_{\boldsymbol{h}^R_2}^{\dag}(t) U_{\boldsymbol{h}^L_2}^{\dag}(t)$.
The way of calculating this integral is the following. First, we move to the path integral representation using Keldysh technique. Then, we shall do an average with respect to Gaussian random variables, and finally, calculate the path integral using the saddle-point approximation.

In order to use Keldysh path integral formalism, we need to construct proper coherent states \cite{Shankar:2017zag}. 
To do so, we introduce an additional set of Majorana operators $\eta_i$ and create fermionic operators in the form:
\begin{equation}
\hat c_i=\frac{\hat \chi_i+i\hat \xi_i}{\sqrt{2}},~~~~\hat c^{\dagger}_i=\frac{\hat \chi_i-i \hat \xi_i}{\sqrt{2}}.
\end{equation}With this notation, the coherent state can be defined as
\begin{equation}
\left| \psi \right>=\left|0 \right>-\psi \left|1 \right>=e^{-\psi \hat{c}^{\dagger} \left|0\right>}, 
\end{equation}
where $\psi$ is a Grassmann variable that parameterizes our state. The set of these states is not orthonormal, and their overlap is given by:
\begin{equation}\label{overl}
\left<\psi | \psi' \right>=\left(\left<0\right|-\left<1\right|\overline{\psi}\right)\left(\left|0\right>-\psi'\left|1\right> \right)=1+\overline{\psi} \psi'=e^{\overline{\psi} \psi'}.
\end{equation}Concerning the path integral formalism, we need to get two more useful identities. First, using the eq.~(\ref{overl}) we can express an identity operator as an integral over the Grasmann fields:
\begin{equation}\label{unit}
\int d \overline{\psi}  \int d \psi e^{-\overline{\psi} \psi} \left|\psi \right> \left< \psi \right|=1,
\end{equation}and secondly we can express the trace of an operator $\hat{O}$ in terms of Grassmann variables as:
\begin{equation} \label{tro}
{\rm Tr}(\hat{O})=\sum_{n_1,n_2,..=0,1}\left<n_1 ...n_N \right|\hat{O} \left| n_1...n_N \right>=\int d \overline{\psi} d\psi e^{-\overline{\psi} \psi} \left<\psi\right|\hat{O} \left|- \psi \right>,
\end{equation}
here $-1$ comes from commuting left and right coherent states. Using eq.~(\ref{unit}) and eq.~(\ref{tro}) we can rewrite eq.~(\ref{framep}) as following
\begin{multline}\label{framecontour}
    F^{(k)}=\frac{1}{\mathcal{N}^{2k}}\int d \eta(h) d \boldsymbol{\overline{\psi}} d\boldsymbol{\psi } \prod_{l=1}^k  \left<\psi^{-,l}_1 \right| U_{1}(\epsilon) \left| \psi^{-,l}_2 \right>\left<\psi^{-,l}_2 \right|U_{2}(\epsilon) \left|\psi^{-,l}_3\right>...\left<\psi^{-,l}_{2N} | -\psi^{-,l}_{1} \right>\cdot\\
\cdot\left<\psi^{+,l}_{2N}\right|U^{\dag}_{2N}(\epsilon) \left| \psi^{+,l}_{2N-1} \right>\left<\psi^{+,l}_{2N-1}\right|U^{\dag}_{2N-1}(\epsilon) \left| \psi^{+,l}_{2N-2} \right>...\left<\psi^{+,l}_{1}| -\psi^{+,l}_{2N} \right>e^{-\overline{\psi}^{s,l}_i \psi^{s,l}_{i} },
\end{multline}where $s=\pm$ indicates one of the two Keldysh contours, $l$ counts replicas, and $i$ labels time. Each matrix element here has the form, 
\begin{equation}
    \left<\psi^{-,l}_1 \right| U_{1}(\epsilon) \left| \psi^{-,l}_2 \right>=\left<\psi^{-,l}_1 \right| e^{ -i^{\frac{q}{2}+1} \epsilon   h_{i_1 i_2..i_q}^1 (\hat c_{i_1}^{\dagger}+\hat c_{i_1})...(\hat c_{i_q}^{\dagger}+\hat c_{i_q})} \left| \psi^{-,l}_2 \right>.
\end{equation}Let us notice that for any matrix element of this form, we can simply replace $\hat c^{\dagger}$ with $\overline{\psi}$ and $\hat c$ with $\psi$. We do not need normal ordering here, as $H$ in eq.~(\ref{sykh}) has one term in which the order is opposite. We can reverse the order at the cost of a minus sign (we stress that $i_k \neq i_l$ everywhere and all fermion operators anti-commute), obtain the action, and once again reverse the order of the corresponding Grassmann variables $\psi, \overline{\psi}$  at the cost of an extra minus sign. Therefore, we shall write the usual Keldysh action: 
\begin{equation}\label{eq:keldyshFermions0}
S(\boldsymbol{\overline{\psi}}, \boldsymbol{\psi})=\sum_{l,s= \pm} s\int_0^T d t \left(  \sum_j \overline{\psi}_j^{s,l} \partial_t \psi_j^{s,l}+H( \overline{\boldsymbol{ \psi}}, \boldsymbol{ \psi})\right).
\end{equation}
 Here the continuous limit was taken $\epsilon \rightarrow 0$, $N \rightarrow \infty$, and $N\epsilon \rightarrow T$. We can now move back to Majorana fermions:
\begin{equation}
\chi^{s,l}_i=\frac{\overline{\psi}^{s,l}_i+\psi^{s,l}_{i}}{\sqrt{2}},~~~~\xi^{s,l}_i=i \frac{\overline{\psi}^{s,l}_i-\psi^{s,l}_{i}}{\sqrt{2}},
\end{equation}note that the variables we consider now are not operators but Grassmann variables and  $i$ is the index of the Fermionic degrees of freedom. In this limit, actions for $\chi$ and $\xi$ fields separate \cite{Shankar:2017zag}, and the partition function factorizes. Therefore, the part that depends on $\eta$ is merely a spectator and can be integrated out eq.~(\ref{framep}) giving a contribution to the overall normalisation.
This can be fixed using the fact that the frame potential must be equal to $1$ at time zero. Therefore, with correct normalisation, we write 
\begin{equation}
     F^{(k)} \equiv \frac{\tilde F^{(k)}(T )}{\tilde F^{(k)}(T = 0)}=\frac{Z(\xi)}{\tilde F^{(k)}(T = 0) \mathcal{N}^{2k}}\int d \eta(h) d \boldsymbol{\chi} e^{-S(\boldsymbol{\chi})},
\end{equation}
where we have
\begin{equation}
    S(\boldsymbol{\chi})=\sum_{l,s= \pm} s\int_0^T d t \left(\frac{1}{2}\sum_j  \chi_j^{s,l} \partial_t \chi_j^{s,l}+i^{\frac{q}{2}+1}\sum_{i_1<i_2<..<i_q} h_{i_1...i_q}(t) \chi^{s,l}_{i_1}... \chi^{s,l}_{i_q} \right),
\end{equation}
and $\tilde F^{(k)}(T)$ is un-normalised value of frame potential. This action is a part of the standard path integral formalism for the SYK model \cite{1806.06840,Gu:2019jub,Lunkin:2020tbq}. Here, we are interested in averaging the Frame potential over the random noisy couplings $h$. To do so, we calculate the Gaussian integral over each of the variables and use the property given by equation (\ref{av}), which defines the variance $\sigma^2$:
\begin{equation}\label{gauss}
    \int d h_{i_1...i_q} e^{-\frac{1}{2 \sigma^2} h^2_{i_1..i_q}+i^{\frac{q}{2}+1} h_{i_1..i_q}M_{i_1..i_q}}= e^{-\frac{(-1)^{\frac{q}{2}}\sigma^2}{2}M^2_{i_1...i_q}},
\end{equation}
where $M_{i_1...i_q}=\sum_{l}(\chi^{-,l}_{i_1}...\chi^{-,l}_{i_q}-\chi^{+,l}_{i_1}...\chi^{+,l}_{i_q})$. Let us introduce the bi-local fields $\hat{G}(t_1,t_2)$, $\hat{\Sigma}(t_1,t_2)$ and perform Hubbard-Stratonovich transformation \footnote{Notice that here in the case of Brownian SYK one could also more simply introduce fields $\hat{G}(t)$ of only one time instead of two, as the interactions are local in time.   }:
\begin{equation}\label{normdelta}
\int d \hat{G} \prod_{s,s'} \delta \left(G^{s s'}_{ll'}(t_1,t_2)-\frac{1}{L} \sum_i \chi^{s,l}_i(t_1) \chi^{s',l'}_i(t_2)\right)= \int d \hat{\Sigma} d \hat{G} e^{- \frac{L}{2} \int d t_1 d t_2 \Sigma^{s s'}_{ll'}\left(G^{s s'}_{ll'}-\frac{1}{L} \chi^{s,l}_i \chi^{s',l'}_i\right)},
\end{equation}
here we suppose the summation over the repeating indexes. Inserting this delta function in the expression for the frame potential we obtain the following expression for an effective action:
\begin{multline}
    S =-\sum_{s,l} s\int_0^T \frac{1}{2} \chi_j^{s,l} \partial_t \chi_j^{s,l}+\frac{L}{4 q^2}\sum_{ l ,l',s,s'} \int_0^T dt_1 dt_2 c_{s s'} \left( 2 G_{l l'}^{s s'}\right)^{q}\delta(t_{12})-\\- \frac{L}{2} \sum_{l,l',s,s'}\int d t_1 d t_2 \Sigma^{s s'}_{ll'}(t_1,t_2)\left(G^{s s'}_{ll'}(t_1,t_2)-\frac{1}{L} \chi^{s,l}_i(t_1) \chi^{s',l'}_i(t_2)\right),
\end{multline}
where
\begin{equation}
    c_{ss'}= - s s'.
\end{equation}
Notice that in order to form bi-local fields in the expression eq.~(\ref{gauss}), we performed $\frac{q(q-1)}{2}$ transpositions of the Grassman variables, which gave us the prefactor in the interaction term $(-1)^{\frac{q(q-1)}{2}}$ together with $(-1)^{\frac{q}{2}}$ from the Gaussian integration, resulting in $(-1)^{\frac{q^2}{2}}$, which is unity due to considering $q$ as an even integer. In order to proceed further, we integrate out the fields $\chi_i$, and the action takes the form:

\begin{align}\label{act}
S=L \log({\rm Pf}(\hat \sigma_z \partial_t -\hat\Sigma))+\frac{L}{4 q^2}\sum\limits_{l,l's,s' } \int_0^T dt_1 dt_2 c_{s s'} \left( 2 G_{l l'}^{s s'}\right)^{q} \delta(t_{12})- \frac{L}{2} \text{Tr}(\hat \Sigma^T \hat G),
\end{align}
where the Pfaffian is taken with respect to indices in time, in Keldysh contours and replica spaces, and the same is applied to the trace. 
Here, the indexes $s$ and $s'$ take two different values, $+$ and $-$, which indicate one of the Keldysh contours. These contours are two loops, as can be seen from equation (\ref{framecontour}). Notice that after averaging over the noise, the fields from the different contours become mixed, and in the action, we have not only $G^{++}$ and $G^{--}$, but also $G^{+-}$ and $G^{-+}$. Therefore, we can think of these operators as matrices in the Keldysh contour $s,s'$ space. Also, let us take into account that the derivative term $\frac{1}{2} \chi \partial_t \chi$ has a positive sign for the $+$ contour and a negative sign for the $-$ contour. Therefore, this term can also be expressed as a $2 \times 2$ matrix with Keldysh contour indexes in the following way:

\begin{equation}
-\sum_{s,l} s\int_0^T \frac{1}{2} \sum_j \chi_j^{s,l} \partial_t \chi_j^{s,l}=-\frac{1}{2}\sum_{l} \sum_j \chi^{l}_j \left( \hat \sigma_z \partial_t \right) \chi^{l}_j.
\end{equation}Here, we introduced the Pauli matrix $\hat\sigma_z$ in the $ss'$ space, and we assume a Kronecker delta in the replica space for this term. In the next section, we are going to calculate this functional integral over the fields $\hat{G}$ and $\hat{\Sigma}$ using the saddle-point approximation at large $L$.

\section{Majorana SYK model and saddle points}\label{sec3}

We are in a position to obtain the equations of motion at the saddle points. To do so, we consider the variation of the action with respect to the bi-local fields:
 \begin{gather}\label{eq:m}
     \frac{\delta S}{\delta \hat{\Sigma}}:~~~   \hat G^{-1}(t_1,t_2)= \hat\sigma_z \partial_t-\hat \hat{\Sigma}(t_1,t_2),~~~~~~~~~~~~~~\\
      \frac{\delta S}{\delta \hat{G}}:~~~   \hat{\Sigma} (t_1,t_2)= \frac{\hat c}{q}  (2\hat{G})^{q-1}(t_1,t_2)\delta(t_{12}),
 \end{gather}
 with the tensor $ \hat c_{s,s'} = c_{ss'}$. 
 In this expression, we suppressed Keldysh and replica indexes, for simplicity.  Setting the variation to zero and summing over all possible solutions, we can find the time dependence of the Frame potential.

\subsection{Short-time behaviour}\label{shorttime}
Let us now proceed naively by looking for a saddle point that is diagonal in the $+-$ space. Then the obvious solution is:
\begin{equation}
    \hat\Sigma=0,~~~~~\hat G^{-1}=\hat \sigma_z \partial_t.
\end{equation}
With this solution, the Frame potential can be expressed as:
\begin{equation}
      F^{(k)}(T=0)=\frac{1}{\mathcal{N}^{2k}} \det(\hat \sigma^z \partial_t)^{\frac{L}{2}} e^{-\frac{L}{4 q^2}\sum\limits_{l l'} \int_0^T d t  (2 G_{l l'}^{++})^q+(2 G_{l l'}^{--})^q}.
 \end{equation}This functional determinant can be calculated using diagonalisation of the operator $\hat \sigma^z \partial_t$. By establishing antiperiodic boundary conditions, we obtain eigenvalues  $\lambda_n=\pm \frac{\pi i (2n+1)}{T}$, where $n=0,\pm 1,\pm 2,..$. In the $+-$ space, we have two eigenvalues $\lambda_n$ and $-\lambda_n$. However, coherent states are over complete, therefore we consider $\lambda_n= \frac{\pi i}{T}, \frac{3 \pi i}{T}, \dots, \frac{(4N+1) \pi i}{4 T}$, which prevents us from treating matrices
 \begin{equation}
\begin{pmatrix}
-\lambda_n & 0\\
0 & \lambda_n 
\end{pmatrix} ~~~\text{and}~~~   \begin{pmatrix}
\lambda_n & 0\\
0 & -\lambda_n 
\end{pmatrix},
\end{equation}
as different matrices. Notice that this determinant should also be taken in the replica space, where we have $k$ copies of the system. Taking the limit $N \rightarrow \infty$ and following the regularization technique from the book \cite{Nakahara:2003nw} (notice that we have $\omega=0$), we get
\begin{equation}
    \det(\hat \sigma_z \partial_t)^{\frac{L}{2}}=2^{\frac{2 k L}{2}},
\end{equation}where $2^{\frac{L}{2}}$ is the dimension of the Hilbert space for the initial operator $U$. Notice that the field $\hat{G}$ is also diagonal in the replica space. Therefore, $G^{++}_{l l'}=\frac{1}{2} \delta_{l l'}=-G_{l l'}^{--}$, and we have
\begin{equation}
    e^{-\frac{L}{4 q^2}\sum_{l l'} \int_0^T d t (2 G_{l l'}^{++})^q+(2 G_{l l'}^{--})^q}=e^{-\frac{LTk}{2 q^2}}.
\end{equation}The Frame potential on this solution takes the form
\begin{equation}
    F^{(k)}= e^{-\frac{LTk}{2 q^2}}.
\end{equation}
We can observe that in this solution, the Frame potential exhibits an exponential decay with respect to time, and all the replicas are factorised. This behaviour implies that the contribution from this part of the Frame potential is relevant during the early times of the system's evolution but becomes increasingly suppressed as time progresses.

\subsection{Long-time behaviour}
In this section, we consider two fundamentally different cases: $q > 2$ and $q = 2$. The difference arises from the fact that in the case of $q = 2$, the action is invariant under the continuous symmetry group $SO(k) \times SO(k)$ (i.e. the replica indices in each Keldysh contour can be rotated separately), therefore the fields $\hat{\Sigma}
$ and $\hat{G}$ can be rotated in the replica space by the generators  $O_1=e^{i \hat A_1 \hat \sigma_z }$, $O_2=e^{i \hat A_2 \hat 1 }$ (which form $SO(k)\times SO(k)$ group), here $\hat A_i$ is a skew-symmetric matrix of the size $k$ and by $\hat A \hat \sigma_z$ we suppose the tensor product. Whereas for $q > 2$, the replica indices are invariant only under the discrete permutation group $P_k$. Therefore, in the first case, the saddle point physics will be characterised by the Goldstone theorem, and the integral over the fluctuations around the saddle point solution will provide us with non-trivial corrections in $L$, which we also aim to study. Another difference between these two examples is that the space of saddle point solutions for $q = 2$ forms a continuous manifold, while for $q > 2$, they are simply a discrete set.

\subsubsection{The non-Gaussian case $q > 2$}
Here, we begin by considering the non-diagonal solution in the $s,s'$ Keldysh contours space. To do so, let us  examine the following ansatz
\begin{equation}\label{ansatzperm}
\hat{G}(t_1,t_2)=\frac{f(|t_{12}|) }{2} \begin{pmatrix}
\Theta(t_{12})-\Theta(-t_{12})&-\hat{\tau}^T\\
\hat{\tau} & -\Theta(t_{12})+\Theta(-t_{12})
\end{pmatrix}.
\end{equation}Here, $f(|t_{12}|)$ is an arbitrary function that depends on the module of difference between time $t_1$ and $t_2$, which will be found later and $\Theta(t_{12})$ is a Heaviside theta function. The matrix $\hat{\tau}$ is a permutation matrix in the replica space. let us  discuss the function ${\rm sgn}(t_{12})$  in the ansatz. In the discretised time one can see that the action consists of the terms $G_{i, i-1}$, where $i$ labels time. As the solution depends on the difference between consequent times $ {\rm sgn}(t_i-t_{i-1})={\rm sgn}(\epsilon \times (i-i+1))={\rm sgn}(\epsilon)$. Therefore,  the ${\rm sgn}(t_{12})$ is never equal to zero in the action, which as we will see later helps us to regularise the action.

The choice of this ansatz is motivated by the fact that we can permute replicas in the action without altering its form. Notice that in the case $q=2$ we have continuous $SO(k) \times SO(k)$ symmetry of the action instead of the permutation symmetry for general $q$. This case will be considered later in this work. By using equation (\ref{eq:m}), we obtain:
\begin{equation}
\hat{\Sigma}(t_1,t_2)= \frac{1}{q} \delta(t_{12}) f(0)^{q-1} \begin{pmatrix}
0&-\hat{\tau}^T\\
\hat{\tau} &0
\end{pmatrix}.
\end{equation}Now, our objective is to determine the function $f(t)$. To achieve this, we perform the Fourier transformation $\hat{\Sigma}(t)=\int d \omega \hat{\Sigma}(\omega) e^{-i \omega t}$ and substitute this ansatz into the first equation of motion (\ref{eq:m}). This procedure gives:
\begin{equation}
\hat G(\omega)=\frac{1 }{\omega^2+\left(\frac{f(0)^{q-1}}{q}\right)^2} \begin{pmatrix}
i \omega&-\frac{f(0)^{q-1}}{q}\hat{\tau}^T\\
\frac{f(0)^{q-1}}{q}\hat{\tau} & -i \omega
\end{pmatrix}.
\end{equation}By performing the inverse Fourier transform and comparing it with the ansatz, we arrive at the following result:
\begin{align}
    \hat{G}(t_1,t_2)=\frac{e^{-\frac{|t_{12}|}{q}} }{2} \begin{pmatrix}
\Theta(t_{12})-\Theta(-t_{12})&-\hat{\tau}^T\\
\hat{\tau} & -\Theta(t_{12})+\Theta(-t_{12})
\end{pmatrix},~~~\hat{\Sigma}(t_1,t_2)= \frac{1}{q} \delta(t_{12}) \begin{pmatrix}
0&-\hat{\tau}^T\\
\hat{\tau} &0
\end{pmatrix}
.\end{align}

Our objective is to understand the behaviour of the action on this solution. Firstly, let us  note that the second term in the action, eq.~(\ref{act}), cancels due to the $c_{s s'}$ prefactor 
\begin{equation}
    \frac{L}{4 q^2}\sum_{ l ,l' s ,s'} \int_0^T dt_1 d t_2 c_{s s'} \left( 2 G_{l l'}^{s s'}\right)^{q}\delta(t_{12})=0.
\end{equation}
On the other hand, the third term will give us a non-trivial contribution 
\begin{equation}
    - \frac{L}{2}\sum\limits_{l,l's,s' } \int_0^T dt_1 d t_2 \Sigma^{s s'}_{ll'}(t_1,t_2)G^{s s'}_{ll'}(t_1,t_2)=-\frac{L}{2} {\rm Tr}(\hat \Sigma^T \hat G)=-\frac{L T k}{2 q}.
\end{equation}
Finally, let us proceed with the calculation of the first term in the action. In fact, the Pfaffian can be understood as a fermionic (quadratic) thermal partition function with the  real inverse temperature equal to $T$:
\begin{equation} \label{pf}
    e^{S_1}=({\rm Pf}[\hat \sigma_z \partial_t-\hat\Sigma])^L={\rm Tr}(e^{-T \hat H})^{L},
\end{equation}
where 
\begin{equation}
    \hat H= -\frac{1}{2}\begin{pmatrix}\hat \chi^+&i \hat \chi^-
        
    \end{pmatrix} \hat \Sigma_0 \begin{pmatrix}
        \hat \chi^+\\
        i \hat \chi^-
    \end{pmatrix},
\end{equation}
and
\begin{equation}
\hat{\Sigma}_0= \frac{1}{q} \begin{pmatrix}
0&-\hat{\tau}^T\\
\hat{\tau} &0
\end{pmatrix}.
\end{equation}
Let us explain the form of the Hamiltonian.  The derivative in the $--$ sector eq.~(\ref{pf}) comes with an extra minus, therefore the Majorana fields in the partition function should come with an extra $i$.  Finally, we obtain
\begin{equation}
    ({\rm Pf}[\hat \sigma_z \partial_t-\hat\Sigma])^L={\rm Tr}(e^{i \frac{T}{q} \hat\chi^- \hat{\tau} \hat\chi^+})^{L}.
\end{equation}
A similar expression was calculated in \cite{ReplicesSymmetry}, and here we use the analogous method. Let us  observe that any arbitrary permutation can be expressed as a direct sum of disjoint cycles, $\hat{\tau}=\bigoplus_i \hat{\tau}^c_i$ which implies, 
\begin{equation}
    \hat{\tau}=
\left(\begin{array}{c|c|c|c}
  \hat{\tau}^c_1
  & 0 & 0&0 \\
\hline
  0 &
  \hat{\tau}^c_2& 0&0\\
  \hline
  0 &
0 &  ...&0
\\
  \hline
  0 &
0 & 0& \hat{\tau}^c_m
\end{array}\right),
\end{equation}then ${\rm Tr}(e^{i \frac{T}{q}  \hat\chi^- \hat{\tau} \hat\chi^+})={\rm Tr}(e^{i \frac{T}{q} \sum_i \hat\chi_i^- \hat{\tau}^c_i \hat\chi_i^+}) $ , where the sum goes over the cycles.  Also, we notice that each cycle can be represented in the canonical form, therefore we have 
\begin{equation}\label{cyc}
    \hat{\tau}^c_i= \begin{cases}
    \delta^{\alpha+1, \beta},~~~n_c ~~~\text{odd}\\
    {\rm {\rm sgn}}(\beta-\alpha)\delta^{\alpha+1,\beta},~~~n_c ~~~\text{even},
    \end{cases}
\end{equation} where $n_c$ represents the length of a cycle, and $\delta^{n_c+1,\beta} = \delta^{1,\beta}$. Here, ${\rm sgn}(\beta-\alpha)$ merely affects the permutation on the boundaries. The boundary conditions have been chosen in a way to keep the parity $P$ equal to $1$. It can be observed that for any $\hat{\tau}$, this solution does not respect the parity of the initial state (the initial state is a thermofield double state between, where for each site, the Fermi parity is $P = 1$). Therefore, the boundary conditions eq.~(\ref{cyc}) arise from this fact. Then, for each cycle, we have:
\begin{equation}
    -T \hat H_c=i \frac{T}{q} \left(\hat \chi^-_{\alpha} \delta^{\alpha+1,\beta} \hat \chi^+_{\beta}+(-1)^{n_c+1} \hat \chi^-_{n_c}\hat \chi^+_1 \right).
\end{equation}
Introducing complex fermions \begin{equation}\label{ferm}
    c_{\alpha}=\frac{\hat \chi^+_{\alpha}+i \hat \chi^-_{\alpha}}{\sqrt{2}}, ~~~~  c^{\dagger}_{\alpha}=\frac{\hat \chi^+_{\alpha}-i \hat \chi^-_{\alpha}}{\sqrt{2}},
\end{equation}
and inserting this into the Hamiltonian we obtain
\begin{equation}
    -T \hat H(\hat{\tau}_c)=\frac{T }{2q} \sum_{\alpha} (c^{\dagger}_{\alpha} c_{\alpha+1}+c_{\alpha} c_{\alpha+1}+h.c.)+\frac{T }{2q}  ((-1)^{n_c+1}c^{\dagger}_n c_1 +h.c.), 
\end{equation} this is the Kitaev chain \cite{Kitaev2001-da} with periodic/antiperiodic boundary conditions, eq.~(\ref{cyc}) and odd/even length of chain respectively. The diagonalisation of the chain is readerly done, giving 
\begin{equation}
    {\rm Tr}(e^{-T \hat H(\hat{\tau}_c)})\sim e^{\frac{n_c T }{2 q}},~~~ T \rightarrow \infty.
\end{equation}
Now we can finally sum over all the cycles, noticing that the size of the permutation matrix $k$ should be equal to the sum of the length of cycles of this permutation $k=\sum_c n_c$ 
\begin{equation}
    {\rm Tr}(e^{-T \hat H})=\prod_c e^{\frac{n_c T }{2 q}}= e^{\frac{k T }{2 q}}.
\end{equation}Consequently we find the following expression for the Pfaffian:
\begin{equation}
    e^{S_1}=({\rm Pf}[\hat \sigma_z \partial_t-\Sigma])^L= e^{\frac{L k T }{2 q}},
\end{equation}
and the Frame potential of this solution can be expressed as:
\begin{equation}
    F^{(k)}(\hat{\tau})=\frac{1}{\mathcal{N}^{2 k}}  e^{S_1+S_2+S_3} =\frac{1}{\mathcal{N}^{2 k}}  e^{\frac{LkT}{2 q}-\frac{LkT}{2 q}}=\frac{1}{\mathcal{N}^{2 k}}.
\end{equation}
We shall now perform the summation of all the possible saddle points, which is the space of permutation $\hat{\tau}$ of size $k$. Moreover, we notice that we are selecting only permutations of a given parity, fixed by the initial state. Therefore, we find, for large $L$,  
\begin{equation}
    F^{(k)} = \sum_{\hat{\tau}}  F^{(k)}(\hat{\tau})   = \frac{k!}{\mathcal{N}^{2 k}} ,
\end{equation}
{namely the same value of the Frame potential for haar-distributed density matrices (with fixed parity).} We have then shown here that the late-time solution converges to the Haar distribution. 
\subsubsection{The Gaussian case $q=2$}
In the case where $q=2$, the action has the continuous symmetry, $SO(k) \times SO(k)$ as discussed at the beginning of the section. Therefore, the ansatz that we examine here consists of orthogonal matrices instead of permutations. Using the same procedure as in the previous subsection, one can see that for $q=2$, the solution is:
\begin{align}
    \hat{G}(t_1,t_2)=\frac{e^{-\frac{|t_{12}|}{q}} }{2} \begin{pmatrix}
\text{{\rm sgn}}(t_{12})&-\hat{\theta}^T\\
\hat{\theta} & -\text{{\rm sgn}}(t_{12})
\end{pmatrix}, ~~~\hat{\Sigma}(t_1,t_2)= \frac{1}{q} \delta(t_{12}) \begin{pmatrix}
0&-\hat{\theta}^T\\
\hat{\theta} &0
\end{pmatrix},
\end{align}where $\hat{\theta}$ is now an orthogonal matrix in the replica space and not a permutation matrix as in the previous case. Again, we are interested in the calculation of the action on this solution. Notice that the second and third terms in the action are the same as in for permutation matrices. Therefore, our objective now is simply to extend the calculation to the first term, namely the Pfaffian. Let us then apply the same procedure to the orthogonal matrices. Any arbitrary orthogonal matrix can be expressed in block diagonal form:
\begin{align}
        \hat{\theta}_{2m}=
\left(\begin{array}{c|c|c|c|c}
  \hat R_1
  & 0 & 0&0&0 \\
\hline
  0 &
  \hat R_2& 0&0&0\\
  \hline 
  0&0&...&0&0\\
  \hline
  0 &
0 &  0&\hat R_{m-1}&0
\\
  \hline
  0 &
0 & 0& 0&\hat R_m
\end{array}\right), ~~~\text{and} ~~~    \hat{\theta}_{2m+1}=
\left(\begin{array}{c|c|c|c|c}
  \hat R_1
  & 0 & 0&0&0 \\
\hline
  0 &
  \hat R_2& 0&0&0\\
  \hline
  0 &
0 &  ...&0&0\\ \hline
  0 &
0 & 0& \hat R_m&0\\
\hline
  0 &
0 & 0& 0&\pm \hat 1\\
\end{array}\right),
\end{align}where, $m$ is a positive integer, and $\hat R_i$ is an orthogonal matrix in two-dimensional space. Then, ${\rm Tr}(e^{i \frac{T}{q} \hat\chi^- \hat{\theta} \hat\chi^+})={\rm Tr}(e^{i \frac{T}{q} \sum_i \hat\chi_i^- \hat R_i \hat\chi_i^+})$. Here, we should also take into account that not all orthogonal matrices will preserve the parity of the initial state. Therefore,  $\hat R_i$ should be rotation matrices. Also, in the case of an odd size, the last element on the diagonal is positive unity as we want to preserve parity.
Now, let us  consider the Hamiltonian for the two-dimensional orthogonal sub-matrix:
\begin{equation}
     - T \hat H_R=i \frac{T}{q}\left( \cos(\phi) \hat \chi^-_1 \hat \chi^+_1+\sin(\phi) \hat\chi^-_1 \hat\chi^+_2-\sin(\phi)\hat \chi^-_2 \hat \chi^+_1+\cos(\phi) \hat \chi^-_2 \hat \chi^+_2\right).
 \end{equation}Here $\phi$ is an angle that parametrizes the rotational matrix, and we can again move to the complex fermions, see eq.~(\ref{ferm}), which gives a simple two-fermions Hamiltonian:
\begin{multline}
        - T \hat H_R=\frac{T}{2 q}(2 \cos(\phi)(c_1^{\dagger}c_1+c_2^{\dagger}c_2-1)+ 2 \sin(\phi)(c_1 c_2-c_1^{\dagger} c_2^{\dagger}))=\\=\frac{T}{ q}\left(c_1^{\dagger}~~c_2 \right)\begin{pmatrix}
\cos{\phi} & - \sin{\phi} \\
- \sin{\phi} & -\cos{\phi}
\end{pmatrix} \begin{pmatrix}
c_1 \\
c^{\dagger}_2 
\end{pmatrix},\end{multline}after the diagonalisation the trace can be easily calculated. Using the well-known formula for the trace of quadratic density matrices $\rm Tr(e^{c^{\dagger}_i \Gamma_{i j} c_j})=\det(1+e^{\Gamma_{ij}})$, we obtain:
\begin{multline}
        {\rm Tr}(e^{-T \hat H_R})=\det\left(1+\exp\left(\frac{T}{q}\begin{pmatrix}
1 & 0 \\
0 & -1 
\end{pmatrix}  \right) \right)=2+e^{\frac{T}{q}}+e^{-\frac{T}{q}}    \rightarrow e^{\frac{T}{q}}, ~~T\rightarrow \infty.
\end{multline}Therefore the Frame potential on the particular solution which depends on an orthogonal matrix $\hat{\theta}$ is
\begin{equation}
    F^{(k)}(\hat{\theta})=\frac{1}{\mathcal{N}^{2 k}}  e^{\frac{LkT}{2 q}-\frac{LkT}{2 q}}=\frac{1}{\mathcal{N}^{2 k}}. 
\end{equation}
As in the previous subsection, the saddle point solution with a given rotational matrix does not depend on the latter. Therefore, we should first consider the fluctuations around the saddle point, and sum over all possible solutions given by the space of orthogonal matrices. As we shall see in the coming section, the zero-mass fluctuations are responsible for different finite-size effects between the $q>2$ and the $q=2$ case. In the $q=2$ case, they carry  $\frac{\log(L)}{L}$ corrections, while with $q>2$ there are no zero-mass modes that can fluctuate around the saddle point solution.  

\subsubsection{Symmetries and Goldstone modes}

In the next subsection, we shall calculate fluctuations around the saddle point, which we discussed earlier. For the case $q=2$, we need to understand how many modes will be massive and carry non-trivial corrections to the Frame potential. To do so, we can use the Goldstone theorem.  Let $B$ be a continuous group of a global symmetry of the action and $H$ is a subgroup of $B$ which leaves the solution of the equations of motion unchanged. Then the number of massless modes is equal to $R_{ml}={\rm dim}(B)-{\rm dim}(H)$. Therefore, the number of massive modes is $R_m=d-R_{ml}$ where $d$ is the number of degrees of freedom of our system. Let us start with the identification of the degrees of freedom. By the definition: 
\begin{equation}
    G_{l l'}=\begin{pmatrix}
        G^{++}_{l l'} & G^{+-}_{l l'}\\
        G^{-+}_{l l'} & G^{--}_{l l'}
    \end{pmatrix},
\end{equation}
and $G^{++}_{l l'}=\frac{1}{L} \sum^L_{i=1} (\chi^+)^i_l (\chi^{+})^i_{l'}$, where $L$ is the number of fermionic modes, first we notice that the matrix $G_{l l'}$ is a skew-symmetric matrix due to the anticommutation relations of the field $\chi$, which contains $2k(2k-1)/2$ degrees of freedom. Let us check the commutation relations for this matrix
\begin{equation}
    [G_{l l'}, G_{m m'}]=\frac{1}{L^2} \sum_i \sum_j [\chi^i_l \chi^i_{l'},\chi^j_m \chi^j_{m'}],
\end{equation}expanding the commutator and using anticommutation relations $\{\chi^i_l,\chi^j_m \}=\frac{1}{2} \delta^{i j} \delta^{l m}$ we have, 
\begin{equation}
    [G_{l l'}, G_{m m'}]=\frac{1}{2 L}(\delta_{l' m}G_{l m'}-\delta_{l m'}G_{m l'}+\delta_{l m} G_{m' l'}-\delta_{l' m'}G_{l m}),
\end{equation}
which gives the commutation relations of rotation generators in $2k$ dimensions. Therefore,  we see that by the definition our matrix $G_{l l'}$ is a generator of $so(2 k)$ algebra. In \cite{Fava:2023tgg}, it was shown that  $S_{l l'}=i G_{l l'}$ satisfy orthogonality relation in the large $L$ limit:
\begin{equation}
    S S^T=\mathbb{I},
\end{equation}
which for the matrix $\hat{G}$ this implies
\begin{equation}
    \hat{G} \hat{G}^T=\mathbb{I}.
\end{equation}
Therefore this matrix should be not only skew-symmetric but also orthogonal. The space of these matrices isomorphic to the $SO(2k )/U(k)$ \cite{baker}, which gives us $d=\frac{2k(2k-1)}{2}-k^2=k(k-1)$ degrees of freedom.  

Now let us consider the group of the global symmetry of the action in more detail. Remind that at finite times, the action looks like eq.~(\ref{act}). Therefore,  after the integration over the fields $\hat{G}$ we have
\begin{equation}
    S = L \log({\rm Pf}(\hat \sigma_z \partial_t -\hat\Sigma))-\frac{L}{4 }\sum\limits_{l,l's,s' } \int_0^{T} dt_1 d t_2 c_{s s'}\left(\Sigma^{s s'}_{l l'} \right)^2 \delta(t_{12}).
\end{equation}
One can see that the problematic term here is $L \log({\rm Pf}(\hat \sigma_z \partial_t -\hat\Sigma))$, which as we already noticed has a rotation symmetry in replica space $O_1=e^{i \hat A_1 \hat \sigma_z }$, $O_2=e^{i \hat A_2 \hat 1 }$ with $\hat A_i$  skew-symmetric matrix of the size $k$ and by $\hat A \hat \sigma_z$ we suppose the tensor product. Notice that matrices $O_0=e^{i \hat A}$ form a $SO(k)$ symmetry group, therefore the space of the symmetries is $SO(k) \times SO(k)$. Therefore,  
\begin{equation}
    \det(\hat \sigma_z \partial_t -O_i\hat\Sigma O_i^T)=\det(O_i O_i^T)\det(O_i^{-1} \hat{\sigma_z} \partial_t (O_i^T)^{-1} -\hat\Sigma), 
\end{equation}
since matrices $O_i$ commute with $\hat \sigma_z$, we have
\begin{equation}
    \det(\hat \sigma_z \partial_t -O_i\hat\Sigma O_i^T)= \det(\hat \sigma_z \partial -\hat\Sigma).
\end{equation}
Other possible transformations are not symmetries of the action, therefore the group is $B=SO(k) \times SO(k)$ and ${\rm dim}(B)=k(k-1)$. 
Remind that the saddle point solution has the form
\begin{multline}
    \hat{G}(t_1,t_2)=\frac{e^{-\frac{|t_{12}|}{2}} }{2} \begin{pmatrix}
{\rm sgn}(t_{12})&-\hat{\theta}^T\\
\hat{\theta} & -{\rm sgn}(t_{12})
\end{pmatrix},~~~ \hat{\Sigma}= \frac{1}{2}  \delta(t_{12}) f(0) \begin{pmatrix}
0&-\hat{\theta}^T\\
\hat{\theta} &0
\end{pmatrix},\end{multline}where $\hat{\theta}$ is an orthogonal matrix $k \times k$. From the form of the solution, we see that the group that allows us to transform one solution into another one is  $B/H=SO(k)$, which means that $H=SO(k)$ too. Therefore,  $R_{ml}={\rm dim}(B)-{\rm dim}(H)=k(k-1)/2$. Therefore, we find
\begin{equation}
    R_{m}=d-R_{ml}=k(k-1)-k(k-1)/2=k(k-1)/2,
\end{equation}
giving the number of massive modes, which will allow us to calculate the integral over the relevant fluctuations.
 The saddle point approximation gives us the following expression for the Frame potential
 \begin{equation}\label{framepotsp}
     \tilde F^{(k)}(T) =\frac{1}{ \mathcal{N}^{2 k}}e^{S_0}\left<\int D \delta G D \delta \Sigma e^{\frac{1}{2}S^{(2)}(\delta G, \delta \Sigma)} \right>_{\hat{\theta}},
 \end{equation}where the on-shell action $S_0$ was found in previous chapters and $\left< ...\right>_{\hat \theta}$ means the summation other all the solutions. Now let us  consider the quadratic fluctuation term $S^{(2)}(\delta G, \delta \Sigma)$:
 \begin{equation}\label{fluctact}
     S^{(2)}(\delta G, \delta \Sigma)=\frac{\delta^2 S}{\delta \hat G \delta \hat G} \delta \hat G^2+\frac{\delta^2 S}{\delta \hat \Sigma \delta \hat \Sigma} \delta \hat \Sigma^2+ \frac{\delta^2 S}{\delta \hat G \delta \hat \Sigma} \delta \hat \Sigma \delta \hat G,
 \end{equation}doing variation we can find 
 \begin{gather}
         \frac{\delta^2 S}{\delta \hat G \delta \hat G} \delta \hat G^2=\frac{L}{2} \sum\limits_{l,l's,s' } \int_0^T dt_1 dt_2  c^{s s'} (\delta G^{s s'}_{l l'})^2\delta(t_{12}), \\
             \frac{\delta^2 S}{\delta \hat G \delta \hat \Sigma} \delta \hat \Sigma \delta \hat G=-L\sum\limits_{l,l's,s' } \int d t_1 d t_2 \delta \Sigma^{s s'}_{ll'}(t_1,t_2) \delta G^{s s'}_{ll'}(t_1,t_2).
 \end{gather}The first term here belongs to $S_0$, the second term vanishes due to the equations of motion, and the third term is the quadratic fluctuation that we are looking for. Therefore,
 \begin{equation}
    \frac{\delta^2 S}{\delta \hat \Sigma \delta \hat \Sigma} \delta \hat \Sigma^2=-\frac{L}{2}\int d t_1 d t_2 G^{s s'}_{l l'}(t_1,t_2) \delta  \Sigma^{s' \alpha}_{l' m}(t_1,t_2) G^{\alpha \alpha'}_{m m'}(t_1,t_2)  \delta \Sigma^{\alpha' s}_{m' l}(t_1,t_2), 
 \end{equation}where the field $\hat{G}$ is a solution of the saddle point equation, and we suppose the summation over the repeating indexes.  Integrating out the fluctuations, we obtain:
 \begin{equation}
     \int D \delta \hat{G} D \delta \hat{\Sigma} e^{\frac{1}{2}S^{(2)}(\delta \hat{G}, \delta \hat{\Sigma})}\sim L^{-R_m}.
 \end{equation}
Here $R_m$ is a number of massive modes. We also need to calculate the normalisation term, which is a standard Gaussian integral 
 \begin{multline}\label{eq:norm}
         \tilde F^{(k)}(0)=\int D \hat{G} D \hat{\Sigma} e^{\left[\frac{L}{4 }\sum\limits_{l,l's,s' } \int_0^{T} dt c_{s s'}\left(G^{s s'}_{l l'} \right)^2-\frac{L}{2} \int d t_1 d t_2 \Sigma^{s s'}_{ll'}(t_1,t_2)G^{s s'}_{ll'}(t_1,t_2)\right]}\sim\\    \sim L^{-\frac{d}{2}} \int D \delta \Sigma e^{\frac{L}{4} \sum\limits_{l,l's,s' } \int_0^T dt c^{s s'} ( \Sigma^{s s'}_{l l'})^2(t)}\sim L^{-d},
 \end{multline} and $d$ is a number of degrees of freedom. Putting it all together, we obtain for $q=2$:
 \begin{equation}
     F^{(k)}_{q=2}=\frac{\tilde F^{(k)}(T)}{\tilde F^{(k)}(0)}\sim \frac{ L^{d-R_m}}{\mathcal{N}^{2 k}} \sim \frac{ L^{R_{ml}}}{\mathcal{N}^{2 k}} ,
 \end{equation}
where $R_{ml}=\frac{k(k-1)}{2}$ is a number of massless modes, taking the logarithm of the Frame potential we can get:\begin{equation} 
     \lim_{L \rightarrow \infty}\left(\frac{1}{L} \log F^{(k)}_{q=2}\right)\sim -k \log(2)+\frac{k(k-1)}{2} \frac{1}{L} \log(L),
 \end{equation}which coincides with the result for Gaussian Haar calculation, see Appendix (\ref{Haarresof}).
 Notice that for the generic case of $q$ interaction, the symmetry of the action and the solution is discrete, therefore there are no Goldstone modes in this case. Due to this fact, the integration over the fluctuation must be carried over all the degrees of freedom which are ${\rm dim}(SO(2 k)/U(k))$, also, we have the same degrees of freedom at zero and at finite time. Therefore, the normalization will cancel the fluctuation part from the finite time action. Hence, for the generic $q>2$ case we have 
 \begin{equation}
     F^{(k)}_{q>2}\sim \frac{k!}{\mathcal{N}^{2 k}} ,
 \end{equation}
 together with,
 \begin{equation}\label{frp:qg2} 
     \lim_{L \rightarrow \infty}\left(\frac{1}{L} \log F^{(k)}_{q>2}\right)\sim -k \log(2),
 \end{equation}
 which indeed coincides with the first correction in system size for gHaar distributed Majorana fermions, see sec. \ref{sec:majo-ghaar}.

\section{Dirac SYK model and Kelsdysh saddle points}\label{sec4}
We shall now consider the Brownian evolution given by the complex SYK Hamiltonian
\begin{equation}\label{HamiltDirac}
    H(t)=\sum_{\substack{1\leq i_1\leq i_2...\leq i_{q/2}\leq L\\1\leq i_{q/2+1}\leq ...\leq i_{q}\leq L}} h_{i_1...i_q}\hat c^{\dagger}_{i_1}..\hat c^{\dagger}_{i_{q/2}}\hat c_{i_{q/2+1}}...\hat c_{i_{q}}.
\end{equation}
Here, $\hat c^{\dagger}$ and $\hat c$ are Dirac Fermions, and standard anti-commutation relations are applied $\{\hat c_{k}, \hat c^{\dagger}_{l}\}=\delta_{k l}$. As in the previous chapter, $h_{i_1...i_q}$ are normally distributed random variables, but this time, they are complex and have the variance:
\begin{equation}
    \left<h_{i_1...i_q }(t_1) h^{*}_{i_1...i_q}(t_2) \right>=\frac{2^{q}(q/2)!^2}{q^2 L^{q-1}  } \delta(t_1-t_2).
\end{equation}
We shall then repeat the analysis of the previous chapter.  The model now possesses an extra symmetry compared to the Majorana case, which is the global $U(1)$ charge conservation 
\begin{equation}
    \hat Q = \sum^L_{i=1} \hat c^\dagger_i \hat c_i . 
\end{equation}
We shall then show how this impacts the saddle point solutions at late time, both for $q>2$ and $q=2$. 

\subsection{Keldysh path integral and saddle points}
We again choose the {\rm TFD} state as the initial state, and the object of our interest has the form given in the eq.~(\ref{framep}), which can be expressed as two Keldysh contours eq.~(\ref{framecontour}). The evolution operator is given by $U_i(\epsilon)=e^{-i\epsilon H(t_i)}$, where the Hamiltonian is from the eq.~(\ref{HamiltDirac}). 
Now we want to derive an effective action of the theory. We proceed similarly as in the Majorana case. Therefore,  the fermion density can be written as  $n=\frac{\left<Q \right>}{L}$.  The Frame potential can be expressed as 
\begin{equation}
    F^{(k)} =\frac{1}{\mathcal{N}^{2k}} \underbrace{\Tr\left[\Texp\left( - i \int_0^T dt H(t)\right)\right]
    \Tr\left[\Texp\left( i \int_0^T dt H(t)\right)\right]\ldots}_{k \mbox{ times}}~~.
\end{equation}Since each of the Hamiltonian conserves the total number of particles $\hat Q = \sum_i \hat c^\dag_i \hat c_i$, we can split each trace into smaller traces over each charge sector, namely
\begin{equation}
     \Tr\left[\Texp\left( \pm i \int_0^T dt H(t)\right)\right]=\sum_{Q^{\pm}} \Tr\left[\Texp\left( \pm i \int_0^T dt H(t)\right)P_{Q^{\pm}}\right],
   \end{equation}
with 
\begin{equation}
     P_{Q^\pm}=\int d \mu ^\pm e^{\pm  i L \mu^\pm\int_0^T dt  (\frac{1}{L}\sum_i \hat c^{\dagger}_i \hat c_i- n^\pm )}.
\end{equation}


Namely, for each Keldysh contour and each replica we can introduce a chemical potential $\mu^s_\ell$, and we denote by $\boldsymbol{\mu}$ the whole set of them. 
Writing down the path integral representation for each trace and by averaging over Hamiltonian realisations, we obtain
\begin{equation}\label{eq:FPDirac-pathintegral}
    F^{(k)} =\frac{1}{\mathcal{N}^{2k}}  \int d\eta(\boldsymbol{h}) \sum_{\boldsymbol{Q} }  
    \int d \boldsymbol{\mu} \int \mathcal{D} \boldsymbol{\psi}    \mathcal{D} \bar{\boldsymbol{\psi}} e^{S(\boldsymbol{\overline{\psi}}, \boldsymbol{\psi})} \prod_{s,l} P_s(\mu^s_l),
\end{equation}
with the action $S(\boldsymbol{\overline{\psi}}, \boldsymbol{\psi})$ already introduced in eq. (\eqref{eq:keldyshFermions0}) and 
with the projectors now written as 
\begin{equation}
    P_s(\mu^{s}_l) = e^{  i s  T  L \mu^{s}_l (\bar{\psi}^{s,l}(0^+) \psi^{s',l'}(0) - n^s_l + \delta_{s,+}\delta_{l l'})   }.
\end{equation}
Antiperiodic boundary conditions are enforced to give the expression of traces. By performing the Hubbard-Stratonovich transformation as usual eq.(\ref{normdelta}) and integrating over Grassmann fields, we obtain the action:
\begin{multline}\label{actferm}
        S = L \log(\det( \hat \sigma_z \partial_t -\hat\Sigma+i \hat \sigma_z \hat \mu  ))+\frac{L}{2 q^2}\sum\limits_{l,l's,s' } \int dt_1 dt_2 \delta(t_{12}) c_{s s'} (2 G^{s s'}_{l l'})^{q/2} (-2 G^{s' s}_{l' l})^{q/2}\\    -L \sum_{l ,l' s ,s'}\int d t_1 d t_2 \Sigma^{s s'}_{ll'}(t_1,t_2)G^{s s'}_{ll'}(t_1,t_2)+i L  T \mu^{+}_l (1-n^+_l)-i L  T \mu^{-}_l (-n^{-}_l).
\end{multline}Here $\hat \mu =\mu^{s}_{l} \delta_{ss'} \delta_{l l'} $ is a matrix diagonal in the Keldysh contours space $s,s'$ and in the replica spaces $l,l'$ .  And the fermionic two point function is given by $G^{s s'}_{l l'}(t)=\frac{1}{L}\left<\sum_i \psi^{s}_{i,l}(t) \overline{\psi}^{s'}_{i,l'}(0) \right>$. We are interested in the equations of motion. Repeating the procedure from the section with Majorana fermions, one can find:
\begin{gather}\label{eq:motionDirac}
        -(\hat G^{-1})^T(t_1,t_2)=  \hat\sigma_z \partial_t-\hat \Sigma(t_1,t_2)+i \sigma_z \hat{\mu} \delta(t_{12}), \\    \Sigma^{s s'}(t_1,t_2)=\frac{c_{ss'}}{q} (2 G^{s s'})^{\frac{q}{2}-1}(t_1,t_2)(-2G^{s' s})^{\frac{q}{2}}(t_1,t_2) \delta(t_{12}),\\
        G^{++}_{l,l'}(0^+)= \delta_{l,l'}(1-n^+_l),~~~G_{l,l'}^{--}(0^+)=- \delta_{l,l'} n^-_l. 
\end{gather}
In the first expression, we suppressed Keldysh and replica indexes, and in the second expression, we suppressed only replica indexes. Notice that for Majorana fermions, by definition, matrices $\hat G$ and $\hat \Sigma$ were real-valued. Now, they are allowed to be complex. 
In the case of single replica $k=1$, in the limit $t\rightarrow  0^{\pm}$ the Green function $G^{s s'}$ was found already in \cite{Zhang:2023vpm, Chen:2020bmq} but here we extend to generic replicas. 

The early-time behaviour can also be calculated similarly \ref{shorttime}, such that the corresponding Frame potential is given by 
\begin{equation}
    F^{(k)}(T\rightarrow 0)=e^{-\frac{LTk}{q^2}}. 
\end{equation}
Now let us consider the late-time solution for $q>2$. Notice that the last two equations of motions, eq. (\ref{eq:motionDirac}) define a boundary conditions for $\hat{G}$, fixed by the charge content. Also, notice that we have the sum over all possible values of  $n^{+(-)}$ in the eq. (\ref{eq:FPDirac-pathintegral}). Therefore, let us first consider the terms with $n^+_l=n^-_l=n_l$. In this case, we have a replica diagonal solution, and it is given by charge charge-dependent matrix, see Appendix (\ref{A:SP})
\begin{equation}
\hat{G}_{\rm diag}=  \begin{pmatrix}
f(t_{12})(\Theta(t_{12})(\hat 1-\hat n))-\Theta(-t_{12})\hat n&-f(t_{12})\hat n\\
f(t_{12})(\hat 1-\hat n) & f(t_{12})(\Theta(-t_{12})(\hat 1-\hat n)-\Theta(t_{12})\hat n)
\end{pmatrix},
\end{equation}and 
\begin{equation}
    \hat \Sigma_{\rm diag}=\frac{2^{q-1}}{q}\begin{pmatrix}
        \frac{1}{2} \Gamma(1-2  \hat n)&-\Gamma(1-\hat n) \\
        \hat n\Gamma&\frac{1}{2} \Gamma(1-2 \hat n)
    \end{pmatrix}\delta(t_{12}),
\end{equation}with $f(t_{12})=e^{-\frac{\Gamma |t_{12}|}{2}-i \hat \mu t_{12}}$ and $\hat n={\rm diag}(n_1,n_2,..,n_k)$ is a diagonal matrix in replica space. This matrix consists of the fermion densities from different replicas and $\Gamma=(\hat n(1-\hat n))^{\frac{q}{2}-1}$ is also a diagonal matrix in replica space.
As in the case of Majorana Fermions, other possible solutions of eq. (\ref{eq:motionDirac}) consists of permutation matrices, as these equations have permutation symmetry in the replica space. However, these solutions are valid only for the terms in the eq. (\ref{eq:motionDirac}) which are given by permutation of the charge content $\hat n^{-}=\hat{\tau} \hat n^{+} \hat{\tau}^T$, due to the last two equations in eq. (\ref{eq:motionDirac}). This gives the solution:
\begin{equation}
\hat{G}=  \begin{pmatrix}
f(t_{12})(\Theta(t_{12})(\hat 1-\hat n)-\Theta(-t_{12})\hat n)&- f(t_{12})\hat n\hat{\tau}^T\\
\hat{\tau} f(t_{12})(\hat 1-\hat n) & \hat{\tau} f(t_{12})(\Theta(-t_{12})(\hat 1-\hat n)-\Theta(t_{12})\hat n )\hat{\tau}^T
\end{pmatrix}.
\end{equation}Let us calculate the on-shell action using this solution.  We shall do our calculation for a replica diagonal solution, as the first three terms in the action are invariant under permutations. Therefore,  we can always  cast our solution to the replica diagonal one and get: 
\begin{equation} \label{det}
    e^{S_1}=(\det[\hat \sigma_z \partial_t-\hat \Sigma_d +i\sigma_z \hat\mu \delta(t_{12})])^L={\rm Tr}(e^{-T \hat H})^L,
\end{equation}where
\begin{gather}
    H=(c_+^{\dagger}~~~ c_-^{\dagger}) \begin{pmatrix}
        \hat 1~&0\\
        0~&\hat i
    \end{pmatrix} H_0 \begin{pmatrix}
        \hat 1~&0\\
        0~&\hat i
    \end{pmatrix} \begin{pmatrix}
        c_+\\
        c_-
    \end{pmatrix},\\
    H_0=\begin{pmatrix}
        \frac{2^{q-1}}{2q}\Gamma(1-2  \hat n)-i \hat \mu&-\frac{2^{q-1}}{q}\Gamma(1-\hat n) \\
        \frac{2^{q-1}}{q}\hat n\Gamma&\frac{2^{q-1}}{2q} \Gamma(1-2 \hat n)+i \hat{\mu}
    \end{pmatrix}. 
\end{gather}Then we use the determinant formula 
\begin{equation}
    {\rm Tr}(e^{-TH})=\det \left(1+{\rm exp}\left(
    {-T \begin{pmatrix}
        \hat 1~&0\\
        0~&\hat i
    \end{pmatrix} H_0 \begin{pmatrix}
        \hat 1~&0\\
        0~&\hat i
    \end{pmatrix}}\right)\right).
\end{equation}Diagonalisation of the matrix exponent gives eigenvalues $ e^{\pm\frac{T \Gamma_l}{2q}-i T \mu_l}=e^{\pm\frac{T (n_l(1-n_l))^{\frac{q}{2}}}{2q}- i T \mu_l}$, where $\Gamma_l$ is a matrix element of the diagonal matrix $\Gamma$. Which gives the answer for the determinant calculation:
\begin{equation}
    {\rm det}(\sigma_z \partial_t-\Sigma_d+i\sigma_z \hat\mu \delta(t_{12}))^L\rightarrow e^{\frac{ 2^{q-1} LT\sum_{l=1}^k \Gamma_l}{2q}-iL T \sum_l \mu_l}.
\end{equation}Now let us  move to the second term in the action:\begin{multline}
    S_2=\frac{L}{2q^2}\sum\limits_{l,l's,s' } \int dt_1 dt_2 \delta(t_{12}) c_{s s'} (2 G^{s s'}_{l l'})^{q/2} (-2 G^{s' s}_{l' l})^{q/2}=\\
    =\frac{L}{q^2}\sum_l\left(n_l^{\frac{q}{2}}(1-n_l)^{\frac{q}{2}}-\int dt (\Theta^q (t)(1-n_l)^{\frac{q}{2}}n_l^{\frac{q}{2}}+\Theta(-t)^q n_l^{\frac{q}{2}}(1-n_l)^{\frac{q}{2}})\delta(t)
    \right)=0,
\end{multline}
notice that both side limit for the sum of Heaviside functions is not well-defined, therefore we assume: 
\begin{align}
    \delta(t)& (\Theta(t)a+\Theta(-t)b)=\frac{1}{2}\left(\lim_{t \rightarrow0^+}(\Theta(t)a+\Theta(-t)b)+\lim_{t \rightarrow0^-}(\Theta(t)a+\Theta(-t)b)\right) \nonumber \\& =\frac{1}{2}(a+b).
\end{align}
Therefore, as in the previous case, this term is equal to zero on the solution due to the forward and backward evolution, which comes with different signs.  The third term in the action gives,
\begin{multline}
       S_3=-L\sum_{l,l',s,s'} \int d t_1 d t_2 \Sigma^{s s'}_{ll'}(t_1,t_2)G^{s s'}_{ll'}(t_1,t_2)=\\=-L \int dt_1 dt_2 \left(\Sigma^{++}_{l l'}G^{++}_{l l'}+\Sigma^{--}_{l l'}G^{--}_{l l'}+\Sigma^{+-}_{l l'}G^{+-}_{l l'}+\Sigma^{-+}_{l l'}G^{-+}_{l l'} \right)=\\=-\frac{2^{q-1}}{2q}\sum_l(1-2n_l)^2 \Gamma_l- \frac{2^{q-1}}{q}\sum_l \Gamma_l(1-n_l)n_l=-\frac{ 2^{q-1} LT\sum_{l=1}^k \Gamma_l}{2q}.
\end{multline}
We find therefore the final expression for the Frame potential
\begin{equation}
    F^{(k)}=\sum_{\hat{\tau}} \int  d \boldsymbol{\mu}^+ \sum_{\boldsymbol{Q}^{+}} \frac{e^{S_1+S_2+S_3}}{\mathcal{N}^{2 k}}=\frac{k! L^k}{\mathcal{N}^{2 k}},
\end{equation}
where we used  $\sum_{\boldsymbol{Q}^{+}} = L^k$. We have then recovered the expression obtained by integration over the Haar group in each charge sector, see Appendix \ref{sec:dirac-haar}. 

\subsection{The Gaussian case $q=2$}

Now let us  consider the case $q=2$:
\begin{gather}\label{eq:motionq2}
        -(\hat G^{-1})^T(t_1,t_2)=  \hat\sigma_z \partial_t-\hat \Sigma(t_1,t_2)+i \sigma_z \hat\mu \delta(t_{12}), \\    \Sigma^{s s'}(t_1,t_2)=\frac{c_{s s'}}{2} (-2G^{s' s})(t_1,t_2) \delta(t_{12}),\\
        G^{++}_l(0^+)=1-n^+_l,~~~G^{--}_l(0^+)=-n^-_l.
\end{gather}Similar to the case with $q>2$ we start with a replica diagonal solution with $n^{+}_l=n^{-}_l$ different for each replica:

\begin{equation}
\hat{G}_{\rm diag}=  \begin{pmatrix}
f(t_{12})(\Theta(t_{12})(\hat 1-\hat{n})-\Theta(-t_{12})\hat{n})&-f(t_{12})\hat{n}\\
f(t_{12})(\hat1-\hat{n}) & f(t_{12})(\Theta(-t_{12})(\hat 1-\hat{n})-\Theta(t_{12})\hat{n})
\end{pmatrix},
\end{equation}where $f(t_{12})=e^{-\frac{|t_{12}|\hat 1}{2}-i \hat \mu t_{12}}$, since $\Gamma=\hat 1$ in the case $q=2$ and
\begin{equation}
    \hat \Sigma_{\rm diag}=\begin{pmatrix}
        \frac{1}{2} (\hat1-2  \hat{n})&-(\hat1-\hat{n}) \\
        \hat{n}&\frac{1}{2}(\hat1-2 \hat{n})
    \end{pmatrix}\delta(t_{12}).
\end{equation}
Since our equations of motion have rotation symmetry in replica space we can consider the solution with an arbitrary unitary matrix $\hat{u}$: 
\begin{equation}
    \hat{G}=  \begin{pmatrix}
f(t_{12})(\Theta(t_{12})(\hat1-\hat{n})-\Theta(-t_{12})\hat{n})&- f(t_{12})\hat{n}\hat{u}^{\dagger}\\
\hat{u} f(t_{12})(\hat1-\hat{n}) & \hat{u} f(t_{12})(\Theta(-t_{12})(\hat1-\hat{n})-\Theta(t_{12})\hat{n} )\hat{u}^{\dagger}
\end{pmatrix}.
\end{equation}
However notice that this solution implies boundary conditions $\hat n^-=\hat{u} \hat n^+ \hat{u}^{\dagger}$, but in the sum over all possible charge content, there are no terms that would satisfy these boundary conditions. Therefore, in the $q=2$ case, this form of the solution does not respect boundary conditions from \ref{eq:motionq2}. 
Another possible solution is given by even charge content for each replica $\hat n=n \delta_{ij} = n \hat{1}$. Notice that for this form of the solution, we have a generic rotated solution, with an arbitrary unitary matrix $\hat{u}$. It is clear that boundary conditions are simplified in this case, as  $n^-  \hat{1}=\hat{u} (n^+ \hat{1}) \hat{u}^{\dagger}=n^+ \hat{1}$, and terms of this form are presented in the sum over the charge content. Therefore,  the solution takes the form:
\begin{equation}\label{eq:SPFerm}
\hat{G}=  \begin{pmatrix}
 f(t_{12})\left(\Theta(t_{12})(\hat 1-n \hat{1})-\Theta(-t_{12})n \hat{1}\right)&- f(t_{12})\hat{u} n \hat{1}\\
f(t_{12})(\hat1-n \hat{1}) \hat{u}^{\dagger} & f(t_{12})(\Theta(-t_{12})(\hat1-n \hat{1})-\Theta(t_{12})n \hat{1})
\end{pmatrix},
\end{equation}
where $f(t_{12})=e^{-\frac{|t_{12}|\hat 1}{2}-i \mu \hat 1 t_{12}}$, with $\mu $ being just a number here. Here $\hat{u}$ is a unitary matrix in $k \times k $ space. Notice also that not all possible matrices $\hat u$ respect the parity of the initial state. Here we need to assume that ${\rm det}(\hat u)=1$, in order to take into account the parity. Therefore, $\hat u $ matrix belongs to $SU(k)$. As in the case of Majorana fermions, the difference between the two cases $q>2$ and $q=2$ will come from the consideration of the fluctuations around the saddle point solution and the presence of the Goldstone modes. The calculation of the action gives the same results as in the previous section and can be done the same way replacing $\hat n \rightarrow n\hat 1$.

\subsubsection{Symmetries, and fluctuations}

Let us start with the consideration of the Goldstone theorem for the case $q=2$. The vector fields $\psi$ and $\overline{\psi}$ have $2k$ components, therefore the group acting in this space is $U(2k)$, with $d=4k^2$ generators. This is the number of degrees of freedom for our system.
Now let us consider the group of the global symmetry of the action. The symmetries of the action are restricted by the term $L \log(\det(\hat{\sigma}_z \partial -\hat{\Sigma}+i \hat \sigma_z \hat \mu))$ and other terms with chemical potential. Notice that in the case $\hat n^+=\hat n^-=n \hat{1}$ and $\hat \mu^+=\hat \mu^-=\mu \hat 1$ the action again has rotation symmetries in the replica space $U_1=e^{i \hat{A}_1 \hat{\sigma}_z}$, $U_2=e^{i \hat{A}_2 \hat{1}}$, where $\hat{A}_i$ is a Hermitian matrix of size $k$. By $\hat{A} \hat{\sigma}_z$, we again mean the tensor product. Notice that matrices $U_0=e^{i \hat{A}}$ form the $U(k)$ group. Therefore, the group of symmetries of the action is $B=U(k) \times U(k)$.

Remember that the saddle point solution has the form eq.~(\ref{eq:SPFerm}) where $\hat{u}$ is a unitary matrix. From the form of the solution, we see that the group that allows us to transform one solution into another is $B/H=SU(k)$. Therefore,  the number of massless modes is $R_{ml}= \text{dim}(B) - \text{dim}(H) = k^2-1$. Then we have the number of massive modes:
\begin{equation}
    R_{m}=d-R_{ml}=4k^2-k^2+1=3k^2.
\end{equation}
Let us consider fluctuations in our system. To do so, we again expand the effective action around the saddle point solution $S=S_0(G_0,\Sigma_0)+S^{(2)}(\delta \hat{G}, \delta \hat{\Sigma})$ and integrate out the fluctuations, which are expected to give us the next order correction in the fermionic degrees of freedom.  The integration over fluctuation in the highest order in $L$ gives us 
\begin{equation}
    \tilde F^{(k)}(T) \sim\sum_{Q=1}^L \frac{1}{ \mathcal{N}^{2 k}} L^{-R_m}.
\end{equation}The normalisation term can be calculated the same way as eq.(\ref{eq:norm}). Which gives the scaling with the system size:
\begin{equation}
       \tilde F^{(k)}(0)\sim L^{-d}.
\end{equation}
Therefore in the final expression for the Frame potential, we get:
\begin{equation}
    F^{(k)}_{q=2}= \frac{\tilde F^{(k)}(T)}{\tilde F^{(k)}(0)}\sim \frac{L }{ \mathcal{N}^{2 k}} L^{R_{ml}},
\end{equation} and the logarithm is
\begin{equation} 
    \lim_{L \rightarrow \infty}\left(\frac{1}{L} \log F^{(k)}_{q=2} \right)\sim -2k \log(2)+k^2 \frac{1}{L} \log(L)+O\left(\frac{1}{L}\right),
\end{equation}which coincides with the result for Gaussian Haar calculation, eq. (\ref{Haarresferm}). As in the case of Majorana fermions, for the generic case of $q$ interactions, there are no Goldstone modes. Due to this fact, the integration over the fluctuations extends over all the degrees of freedom, which are $ \text{dim}(U(2k))$. This coincides with the number of degrees of freedom at both zero and finite time. Therefore, the normalization will cancel the fluctuation part from the finite-time action. Hence, for the generic case of $q>2$, we have:
\begin{equation} 
    \lim_{L \rightarrow \infty}\left(\frac{1}{L} \log F^{(k)}_{q>2}\right)\sim -2 k \log(2)+O\left(\frac{1}{L} \right).
\end{equation}Notice that this result also coincides with the gHaar integrated result in eq.~(\ref{Haaru}).

\section{Conclusion and discussion}\label{sec5}

In conclusion, this paper explores the degree of quantum mixing given by the unitary time evolution of the Brownian SYK model, which serves as a paradigmatic model for quantum scrambling and chaos. Our investigation focuses on the Frame potential, a direct quantifier of $k$-th design realisation. By considering two identical copies of an initial density matrix and calculating the overlap between their independent time evolutions, the Frame potential, denoted as $F^{(k)}(T)$, is evaluated. From a study of the short and late-time saddle points of the associated Keldysh action, we find that the Frame potential exhibits exponential decay at early times and saturates to an exponentially small, with the size of the system value at a timescale only dependent on the number $q$ of interacting fermions.  Our analysis reveals that in the case of non-Gaussian SYK models with $q>2$, the Frame potential converges to its Haar-averaged value, denoted as $F_{\rm Haar}$, at late times. Such late-time behaviour manifests that the systems fully explore the available Hilbert space in a uniform manner. However, a difference arises in the Gaussian integrable case with two-fermion interactions. In this scenario, the evolution exhibits mixing behaviour solely within the space of Gaussian states. Here, a Gaussian Haar (gHaar) measure is introduced, and it is demonstrated that the Frame potential converges to the gHaar measure at late times, associated with the value of the Frame potential denoted as $F_{\rm Haar}$. Remarkably, the value of the gHaar-integrated Frame potential is equivalent to the generic case in the limit of a large fermion number $L$, albeit with logarithmic corrections in $L$. The presence of these corrections arises from Goldstone fluctuations around the late-time saddle point. Notably, these fluctuations are only present in the Gaussian case, where both the Keldysh action and the saddle point have a continuous group symmetry. In the complex (Dirac) SYK case, the total charge conservation forces the time evolution to reach Haar only within each charge sector. We have shown that by introducing projectors over the different charge sectors, one can introduce saddle point equations with different charge content between different replicas. While in the interacting $q>2$ case the solution with different charges dominates, in the $q=2$ case instead the one with all the replicas in the same sector dominates, as the latter possesses the largest number of massless modes. Overall, as expected, the constraint of the global charge conservation leads to values of the Frame potential at large times which are larger than the corresponding ones in the Majorana case.

The study of scrambling dynamics in the context of continuous monitoring has received significant attention in recent years, with a particular focus on measurement-induced phase transitions in fermionic systems \cite{2108.11973,2302.12820,PhysRevX.9.031009,Antonini2022}. Exploring for example the effects of continuous monitoring on the Frame potential in the SYK model represents a promising avenue for extending the current work.  Additionally, investigating the behaviour of the Frame potential in non-Brownian dynamics for the SYK model presents another compelling extension. 

\section*{Acknowledgement} We thank Lorenzo Piroli, Guido Giachetti, and Adam Nahum for useful discussions. This work has been partially funded by the ERC Starting Grant 101042293 (HEPIQ) (J.D.N. and A.T.) and by the ANR JCJC grant ANR-21-CE47-0003 (TamEnt) (A.D.L.).

\appendix
\section{Haar-averaged Frame potential}
As we discussed during our work, it is expected for the Frame potential to converge to its Haar value during Brownian evolution. In this appendix, we calculate the Haar values for the Frame potential.
First, we consider the case when the evolution matrix $U$ is selected from the corresponding group using the Haar measure. The groups we consider here are the orthogonal group for Majorana fermions and the unitary group for Dirac fermions. These Haar values correspond to the evolution with an arbitrary number of interacting fermions ($q>2$).
Second, we consider the case where the evolution matrix is quadratic in fermions ($q=2$), i.e., $U=e^{c^{\dagger}h c}$. In this case, the matrix values are again chosen with respect to the Haar measure, but the space of integration is smaller compared to the full Haar average.



\subsection{Gaussian-Haar averages for Majorana fermions}\label{sec:majo-ghaar}
If we consider Gaussian evolution, then its Haar distribution is the unitary evolution is a Gaussian $U$, i.e. $U=e^{\frac{1}{2} h_{i j} \chi_i \chi_j}$, where $\chi_j$ are Majorana fields, then we can express  ${\rm Tr}(U)={\rm Pf}(1+e^{h_{i j}})$ (as for example shown in \cite{Klich2014ANO})  where $e^{h_{i j}}=\hat{u}$ is an orthogonal matrix with the size $L\times L$  ($L$ is even). We are interested in the case when matrices $\hat{u}$ are random with respect to the Haar measure. Notice that we are averaging over matrices with $\det(\hat{u})=1$, as $\hat{u}=e^{\frac{1}{2}h}$ where $h$ is an antisymmetric matrix (all the eigenvalues of $h$ have conjugated pairs and lie on the imaginary axis, it means that eigenvalues of $\hat{u}$ are $\lambda_1=e^{i b_1}, \lambda^*_1=e^{-i b_1},...$ and their product is 1). This gives us the following averaging
\begin{multline}
         F^k _{\rm gHaar}=\frac{1}{\mathcal{N}^{2k}} \int \left|\det(1+\hat{u}) \right|^{ k} d \mu(\hat{u})=\\    =\frac{1}{\mathcal{N}^{2k}} \int \left|\prod^L_{j=1} (1+e^{i \theta_j}) \right|^{k} \prod_{i< j} \left| 2(\cos \theta_i-\cos \theta_j ) \right|^2 \frac{d \theta_1..d \theta_{L/2}}{\left(\frac{L}{2} \right)! (2 \pi)^{\frac{L}{2}}}=\\ =\frac{1}{\mathcal{N}^{2k}} \int \left|\prod^{L/2}_{j=1} (2+2 \cos(\theta_j)) \right|^{k} \prod_{i< j} \left| 2(\cos \theta_i-\cos \theta_j ) \right|^2 \frac{d \theta_1..d \theta_{L/2}}{\left(\frac{L}{2} \right)! (2 \pi)^{\frac{L}{2}}}. 
\end{multline}In the third line, we used the fact that all eigenvalues have their conjugated pair. The measure in this expression comes from the fact that we integrate other orthogonal matrices with the determinant equal to 1 and even size(see chapter 2.6 in \cite{Forrester}).
Doing change of variables $  \cos(\theta)=t$ and calculating the Jacobian of this transformation we can get 
\begin{multline}\label{framehaar}
         F^k _{\rm gHaar}=\frac{1}{\mathcal{N}^{2k}}\frac{1}{F^0}\int_{-1}^1 d t_1 ..\int_{-1}^1dt_{L/2} \prod^{L/2}_{j=1} |2(1+t_j)|^{k}\prod_{i<j} |t_i-t_j|^2 \prod_{k=1}^{L/2} (1-t_k^2)^{-\frac{1}{2}}.
\end{multline}
Now we need to calculate these integrals. For simplicity, we can introduce the function to name these integrals :
\begin{equation}
    F=2^{\frac{Lk}{2}} \int_{-1}^1 d t_1 ..\int_{-1}^1dt_{L/2} \prod^{L/2}_{j=1} |1+t_j|^{k}\prod_{i<j} |t_i-t_j|^2 \prod_{k=1}^{L/2} (1-t_k^2)^{-\frac{1}{2}}.
\end{equation}Here we can notice that it is just a special case of Selberg integral (see chapters $3.6, 3.7 , 4.1, 4.7$ in \cite{Forrester})
\begin{equation}
    F=2^{\frac{Lk}{2}}2^{\frac{L}{2}+\frac{L(k-1)}{2}+\frac{L}{2}(\frac{L}{2}-1)}S_{\frac{L}{2}}\left(k-\frac{1}{2},-\frac{1}{2},1\right)=2^{Lk}2^{\frac{L}{2}(\frac{L}{2}-1)}S_{\frac{L}{2}}\left(k-\frac{1}{2},-\frac{1}{2},1\right).
\end{equation}
The Selberg integral has a well-known expression in terms of Gamma functions, which is:
\begin{multline}
        S_L(\lambda_1,\lambda_2,\lambda)=\int_0^1 d t_1...\int_0^1 d t_L \prod_{l=1}^L t_l^{\lambda_1}(1-t_l)^{\lambda_2} \prod_{k<j}|t_k-t_j|^{2 \lambda}=\\=\prod_{j=0}^{L-1}\frac{\Gamma(\lambda_1+1+j \lambda) \Gamma(\lambda_2+1+j \lambda)\Gamma(1+(j+1)\lambda)}{\Gamma(\lambda_1+\lambda_2+2+(L+j-1)\lambda) \Gamma(\lambda+1)}.
\end{multline}Inserting this solution into the eq.~(\ref{framehaar}) we can have the final expression
\begin{multline}
         F^k _{\rm gHaar}= \frac{1}{\mathcal{N}^{2k}}\frac{2^{Lk}2^{\frac{L}{2}(\frac{L}{2}-1)}S_{\frac{L}{2}}\left(k-\frac{1}{2},-\frac{1}{2},1\right)}{2^{\frac{L}{2}(\frac{L}{2}-1)}S_{\frac{L}{2}}\left(-\frac{1}{2},-\frac{1}{2},1\right)}=\frac{S_{\frac{L}{2}}\left(k-\frac{1}{2},-\frac{1}{2},1\right)}{S_{\frac{L}{2}}\left(-\frac{1}{2},-\frac{1}{2},1\right)}=\\    =\prod_{j=0}^{\frac{L}{2}-1}\frac{\Gamma(k+\frac{1}{2}+j)}{\Gamma(\frac{1}{2}+j)} \frac{\Gamma(\frac{L}{2}+j)}{\Gamma(k+\frac{N}{2}+j)}\sim \frac{c_k L^{k(k-1)/2}}{\mathcal{N}^{2k}}+...,
\end{multline}where $c_k=\frac{2}{2^{\frac{(k-1)(k-2)}{2}}} \prod_{i=0}^{k-1}\frac{1}{(2i-1)!!} $ and we define $(-1)!!=1$. Therefore, for $k=1$ we have
\begin{equation}
     F^1 _{\rm gHaar}=\frac{2}{2^L}= F^1 _{\rm Haar},
\end{equation}
which is equal to the Frame potential in the case when the evolution matrix was taken from the Haar ensemble of Orthogonal matrices. 
We are interested in the large $L$ limit expansion of the Frame potential for arbitrary k, therefore we can use a "Stirling-like"  formula for the Barnes functions, which gives: 
\begin{equation}\label{Haarresof}
    \lim_{L \rightarrow \infty} \left(\frac{1}{L}\log\left( F^k _{\rm gHaar}\right)\right)\sim-k \log(2)+\frac{k(k-1)}{2}\frac{1}{L} \log(L).
\end{equation}

\subsection{Haar average of frame potential for Dirac fermions}\label{sec:dirac-haar}
Now we focus on the Dirac fermions case. In this case, the large time Haar distribution for the evolution operator is split into different charge blocks, where each block is a unitary random matrix. The Frame potential then takes the form  
\begin{equation}
    F^{(k)}_{\rm Haar}=\frac{1}{\mathcal{N}^{2k}} \Big[\prod_Q \int_{\rm Haar}  d U_Q \Big] \left| \sum_Q {\rm Tr}\left[ U_Q\right] \right|^{2 k}.
\end{equation}
Given that only paired traces give a finite result, we obtain $k!$ pairs whose average is always equal to $1$, as
\begin{equation}
      \int_{\rm Haar}   d U \left|   {\rm Tr}\left[ U\right] \right|^{2 }= 1,
\end{equation}
therefore using $\sum_Q = L$, we obtain 
\begin{equation}\label{Haaru}
    \frac{1}{\mathcal{N}^{2k}} \int_{\rm Haar} d U \left|{\rm Tr}\left[ U\right] \right|^{2 k}=\frac{L^k}{\mathcal{N}^{2k}}(k)!.
\end{equation} 

\subsection{Gaussian-Haar averages for Dirac fermions}\label{sec:dirac-ghaar}
In the case of Gaussian evolution, we consider a unitary Gaussian operator $U$, i.e. $U=e^{i h_{i j} c^{\dagger}_i c_j}$, where $c_j$ are Dirac fermions. Then we can express  ${\rm Tr}(U)={\rm det}(1+\hat{u})$ where $e^{i h_{i j}}=\hat{u}$ is a unitary matrix with the size $L\times L$  (with $L$ even). The matrices $\hat{u}$ are again random with respect to the Haar measure. This gives us the following average,
\begin{multline}
      F^{(k)}_{\rm gHaar}=\frac{1}{\mathcal{N}^{2k}} \int \left|\det(1+\hat{u}) \right|^{2 k} d \mu(\hat{u})= \\     =\frac{1}{\mathcal{N}^{2k}} \int \prod^L_{j=1} \left|(1+e^{i \theta_j}) \right|^{ 2k} \prod_{i< j} \left| e^{i \theta_i}-e^{i \theta_j} \right|^2 \frac{d \theta_1..d \theta_N}{N! (2 \pi)^N}.
\end{multline}
The measure in this expression comes from the fact that we integrate other unitary matrices (see chapter 2.6 in \cite{Forrester}).  This integral is a Morris integral which can be expressed in terms of Gamma functions:
\begin{multline}
        M_L(a,b,\lambda)=\int_{-1/2}^{1/2} \prod^L_{j=1} e^{i \theta_i (a-b)/2} \left|(1+e^{i \theta_j}) \right|^{ a+b} \prod_{i< j} \left| e^{i \theta_i}-e^{i \theta_j} \right|^{2\lambda}d \theta_1..d \theta_N=\\=\prod_{j=0}^{L-1}\frac{\Gamma(\lambda j+a+b+1)\Gamma(\lambda(j+1)+1)}{\Gamma(\lambda j +a+1)\Gamma(\lambda j +b+1) \Gamma(1+\lambda)}.
\end{multline}Inserting this into the original expression for the Frame potential we can get:\begin{equation}
     F^k _{\rm gHaar}= \frac{1}{\mathcal{N}^{2k}}  \frac{M_L(k,k,1)}{M_L(0,0,1)}=\frac{1}{\mathcal{N}^{2k}} \prod^{L-1}_{j=0} \frac{\Gamma(j+2k+1)\Gamma(j+1)}{\Gamma(j+k+1)^2}.
\end{equation}Again we are interested in the large $L$ limit expansion:
\begin{equation}\label{Haarresferm}
     F^{(k)}_{\rm Haar} = \frac{\tilde{c}_k L^{k^2}}{\mathcal{N}^{2k}} + \ldots
\end{equation}
where we neglected subleading corrections in $L$,  and $\tilde{c}_k=\frac{{\rm sf}^2(k-1)}{{\rm sf}(2k-1)}$ with ${\rm sf}(n)=1!2!...n!$ a super factorial. Comparing this to the full Haar average, we again find 
\begin{equation}
     F^k _{\rm gHaar}\geq  F^k _{\rm Haar},
\end{equation}therefore the full Haar average is again smaller than the fermionic one.
\section{Dirac SYK saddle point with fixed charge density}\label{A:SP}

Let's start with replica-diagonal solution $Q^{+}_l=Q^-_l$, we use the following ansatz 

\begin{equation}
\hat{G}_d= \frac{1}{2} \begin{pmatrix}
f(\Theta(t_{12})(\hat 1-\hat n)-\Theta(-t_{12})\hat n)&-f\hat n\\
f(\hat 1-\hat n) & f(\Theta(-t_{12})(\hat 1-\hat n)-\Theta(t_{12})\hat n)
\end{pmatrix},
\end{equation}
where  $f_{ll'}(t)$ is a time-dependent matrix in replica space with initial condition $f_{l l'}(0)=f(0) \delta_{l l'}$. First, we consider the equation 
\begin{equation}
    \Sigma^{s s'}_{l l'}(t)=\frac{c_{s s'}}{q} (2 G^{s s'}_{l l'})^{\frac{q}{2}-1}(t)(-2G^{s' s}_{l'l})^{\frac{q}{2}}(-t) \delta(t),
\end{equation}
which gives 
\begin{equation}
    \hat \Sigma_d=\frac{f^{q-1}(0)}{q}\begin{pmatrix}
        \frac{1}{2} \Gamma(\hat 1-2  \hat n)&-\Gamma(\hat 1-\hat n) \\
        \hat n\Gamma&\frac{1}{2} \Gamma(\hat 1-2 \hat n)
    \end{pmatrix}\delta(t_{12}),
\end{equation}
where  $\hat n={\rm diag}(n_1,n_2,..,n_k)$ is a diagonal matrix in replica space, which consists of the fermion densities from different replicas, and $\Gamma=(\hat n(\hat 1-\hat n))^{\frac{q}{2}-1}$ is also a diagonal matrix. Using the equation $-(\hat G^{-1})^{s's}_{l'l}(t)=  \hat\sigma_z \partial_t-\hat \Sigma^{ss'}_{l l'}(t)+i \sigma_z\mu^{s}_l$ and Fourier transformation, we can find 
\begin{equation}
    \hat G^{-1}_d=\begin{pmatrix}
        i \omega+\frac{1}{2} \tilde \Gamma (\hat 1-2   \hat n)-i \hat \mu^+& \tilde \Gamma \hat n \\
        -(\hat 1-\hat n)\tilde\Gamma  &-i \omega+\frac{1}{2}\tilde \Gamma(\hat 1-2 \hat n)+i \hat \mu^-,
    \end{pmatrix}
\end{equation}
where $\tilde \Gamma =\frac{f^{q-1}(0)}{q} \Gamma$. Here we can use the formula for the block matrix and assume $\mu^+_l=\mu^-_l=\mu_l$ (since $Q^{+}_l=Q^-_l$), therefore we find
\begin{equation}
    \begin{pmatrix}
        A&B\\
        C&D
    \end{pmatrix}^{-1}=\begin{pmatrix}
        A^{-1}+A^{-1}B(D-CA^{-1}B)^{-1}CA^{-1}&-A^{-1}B(D-CA^{-1}B)^{-1}\\
        -(D-CA^{-1}B)^{-1}CA^{-1}&(D-CA^{-1}B)^{-1}
    \end{pmatrix},
\end{equation}

which gives 
\begin{equation}
    \hat G_d=\frac{1}{ (\omega-\hat \mu)^2+\tilde \Gamma^2 \frac{1}{4}}\begin{pmatrix}
        -i \omega+\tilde\Gamma/2(1-2\hat n)+i \hat \mu  &-
     \tilde \Gamma \hat n\\
        (1-\hat n)\tilde \Gamma&i \omega+\frac{1}{2}\tilde \Gamma (1-2\hat n)-i \hat \mu
    \end{pmatrix}.
\end{equation}
The inverse Fourier transform of each element of the matrix gives 
\begin{equation}
\hat{G}_d=  \begin{pmatrix}
e^{-\frac{\Gamma |t_{12}|}{2}-i \hat{\mu} t_{12}}(\Theta(t_{12})(\hat 1-\hat n)-\Theta(-t_{12})\hat n)&-e^{-\frac{\Gamma |t_{12}|}{2}-i \hat{\mu} t_{12}}\hat n\\
e^{-\frac{\Gamma |t_{12}|}{2}-i \hat{\mu} t_{12}}(\hat 1-\hat n) & e^{-\frac{\Gamma |t_{12}|}{2}-i \hat{\mu} t_{12}}(\Theta(-t_{12})(\hat 1-\hat n)-\Theta(t_{12})\hat n)
\end{pmatrix}.
\end{equation}

\bibliographystyle{unsrt}
\bibliography{biblio}
\end{document}